\documentclass[useAMS,usenatbib]{mn2e}

\title[Gamma-ray production in young open clusters]
{Gamma-ray production in young open clusters: Berk 87, Cyg OB2 and Westerlund 2}
\author[W. Bednarek]{W. Bednarek\thanks{E-mail:
bednar@fizwe4.phys.uni.lodz.pl}\\
Department of Experimental Physics, University of \L \'od\'z,
ul. Pomorska 149/153, 90-236 \L \'od\'z, Poland}
\begin{document}

\date{Accepted . Received ; in original form }

\pagerange{\pageref{firstpage}--\pageref{lastpage}} \pubyear{2007}

\maketitle

\label{firstpage}

\begin{abstract}
Young open clusters are sites of cosmic ray acceleration as indicated by recent detections of the TeV $\gamma$-ray sources in the directions of two open clusters
(Cyg OB2 and Westerlund 2). In fact, up to now a few different scenarios for acceleration of particles inside open clusters have been considered, i.e. shocks in massive star winds, pulsars and their nebulae, supernova shocks, massive compact binaries. Here we consider in detail the radiation processes due to both electrons and hadrons accelerated inside the open cluster. As a specific scenario, we apply the acceleration process at the shocks arising in the winds of WR type stars. Particles diffuse through the medium of the open cluster during the activity time of the acceleration scenario defined by the age of the WR star. They interact with the matter and radiation, at first inside the open cluster and, later in the dense surrounding clouds. We calculate the broad band spectrum in different processes for three example open clusters (Berk 87, Cyg OB2, Westerlund 2) for which the best observational constraints on the spectra are at present available. It is assumed that the
high energy phenomena, observed from the X-ray up to the GeV-TeV $\gamma$-ray energies, are related to each other. We conclude that the most likely description of the radiation processes in these objects is achieved in the hybrid (leptonic-hadronic) model in which leptons are responsible for the observed X-ray and GeV $\gamma$-ray emission and hadrons are responsible for the TeV $\gamma$-ray emission. 
\end{abstract}

\section{Introduction}

Open clusters are concentrations of young stars surrounded by high density clouds.
With the characteristic age of a few to several million years, they still contain many early type stars (OB and WR type) which produce strong radiation field and energetic winds.
A fraction of the wind energy can be transfered to relativistic particles by diffuse shock acceleration 
process occurring at the shocked wind boundary and turbulent wind itself (see e.g. V\"olk \& Forman~1982, Cesarsky \& Montmerle~1983). Moreover, different type of objects related to the massive star evolution, e.g. supernova remnants, pulsar wind nebulae, massive binary systems,
can also be responsible for acceleration of particles. Therefore, open clusters are expected to be likely sources of high energy neutral radiation ($\gamma$-rays, neutrinos, neutrons) produced in collisions of particles with the matter and soft radiation. 

In fact, some open clusters have been found in the relatively large error boxes of the EGRET unidentified sources reported in the 3rd EGRET catalog (Hartman et al.~1999), e.g. 3EG J2021+4716 and 3EG J2016+3657 - Berk 87, 3EG J2033+4118 - Cyg OB2, or 3EG J1027-5817 - Westerlund 2. The TeV sources have been also reported inside
the Cyg OB2 by the HEGRA group (Aharonian et al.~2002), Westerlund 2 by the HESS group 
(Aharonian et al.~2006), and Berk 87 by the Milagro group (Abdo et al.~2006).
Recently, the upper limits on the TeV fluxes from 14 open clusters have been reported by the 
HEGRA group (Aharonian et al.~2006) and from some unidentified EGRET sources toward the open 
clusters by the Whipple group (Fegan et al.~2005).

A few specific models have been already considered for the $\gamma$-ray production in the open clusters. For example, Giovannelli et al.~(1996) and Aharonian et al.~(2006)
investigate the $\gamma$-ray radiation processes by hadrons accelerated at the shocks in the massive star winds inside the cluster.
Torres at al.~(2004) and Domingo-Santamaria \& Torres~(2006) also apply such general scenario 
arguing that accelerated hadrons can penetrate the winds of massive stars. They are able to 
produce $\gamma$-rays and neutrinos in interactions with dense parts of the wind close to the 
stellar surface. Bednarek~(2003) propose that during the age of the open cluster 
(a few million years) some of the most massive stars should already explode as supernovae creating 
energetic pulsars surrounded by the pulsar wind nebulae. Nuclei accelerated in the vicinity of 
the pulsar (inside the nebula) interact with the surrounding matter and radiation producing 
TeV $\gamma$-rays. On the other hand, pulsars can be responsible for the observed unidentified 
EGRET sources. Also the supernova shock waves can accelerate particles which might produce high 
energy radiation (Butt et al.~2007). If nuclei are efficiently accelerated inside the open 
cluster, then $\gamma$-rays can be also produced in the de-excitation of heavy nuclei 
(Anchordoqui et al.~2006).

In this paper we adopt a scenario for the acceleration of particles inside open clusters 
originally proposed by e.g. Cesarsky \& Montmerle~(1983). The early version of the pure hadronic model has been already considered as responsible for the TeV $\gamma$-ray flux from Berk 87 (e.g. Giovannelli et al.~1996, Aharonian et al. 2006).
In the present paper, we study this general scenario in a more detail assuming that both,
electrons and hadrons, can be accelerated at the shocks formed as a result of interaction of 
strong winds from the Wolf-Rayet (WR) type stars with the matter and radiation inside the open cluster and surrounding dense clouds. We calculate $\gamma$-ray fluxes expected from leptonic and hadronic processes for the range of parameters which describe the conditions inside specific open clusters.

\section{The content of open clusters}

\begin{table*}
\begin{center}
\caption{The open clusters basic parameters and their relation to the gamma-ray sources.}             
\label{tab2}      
\begin{tabular}{c c c c}     
\hline
Parameter      &  Berk 87  &  Cyg OB2  &  Wester 2 \\
\hline
distance       & 0.95 kpc  &  1.7 kpc     & 8 kpc     \\
core radius    & 2 pc      &  14 pc       & 6 pc      \\
massive stars  & WR 142    & 120 O type  & WR 20a    \\
stellar luminosity  & $2\times 10^{39}$ erg s$^{-1}$  & $1.4\times 10^{41}$ erg s$^{-1}$  &  $2\times 10^{40}$ erg s$^{-1}$      \\
energy density   &  260 eV cm$^{-3}$   & 120 eV cm$^{-3}$   &  500 eV cm$^{-3}$ \\
power of winds  &  $10^{38}$ erg s$^{-1}$  &   $1.5\times 10^{38}$ erg s$^{-1}$  & $2\times 10^{38}$ erg s$^{-1}$  \\
EGRET source    & 3EG J2021+4716     &  3EG J2033+4118   &  3EG J1027-5817    \\
                & 3EG J2016+3657     &                   &                    \\
                & GeV J2020+3658     &                   &                    \\
TeV source      & MGRO J2019+37      &  HEGRA J2032+4130 & HESS J1023-575     \\
                &                    &  Milagro          &                    \\
\hline                  
\end{tabular}
\end{center}
\end{table*}

We intend to discus, as an example, three open clusters: Berk 87,
Cyg OB2, and Westerlund2. They are either close or inside the (relatively large) error boxes of the unidentified EGRET sources. That's way, they were also targets for the Cherenkov telescopes at TeV energies. In this section we introduce the basic parameters of these objects available in the literature (see Table~1). All these open clusters are surrounded by high density regions (Dent, MacDonald \& Andersson~1988, Butt et al.~2003), have large stellar luminosity and contains early type stars. We estimate the total stellar luminosities of these objects and, for the known radii, calculate the average energy density of stellar radiation (see Table.~1). 
The winds of the massive stars in these open clusters collide with surrounding matter and between themselves giving rise to the shock waves on which particles can be accelerated to relativistic energies (e.g. Cesarsky \& Montmerle~1983). These particles produce synchrotron X-ray emission which can be observed from some of them.
The effective energy density of stellar photons, which are seen by accelerated electrons, is likely to be larger since at least a part of the electrons might diffuse against the massive star winds being exposed to much denser effective stellar radiation field (see e.g. Torres et al.~2004). We take this possibility into account by introducing the enhancement factor of the stellar radiation, $\mu = U_{\rm rad}^{eff}/U_{\rm rad}$, describing the effective energy density of stellar radiation with which electrons can interact. This factor depends on the details of the diffusion and advection processes of electrons in the vicinity of specific stars of the open cluster. These processes are not well understood since they depend on the wind and magnetic field models for specific stars present in the open clusters.

\section{Acceleration of particles inside open clusters} 

We assume that a part, $\eta_{\rm e}$, of the WR star wind energy can go on acceleration of electrons and a part, $\eta_{\rm p}$, go on acceleration of hadrons. These coefficients are kept as free parameters, although it is believed, based on the required efficiency of acceleration of the cosmic rays in the Galaxy by the supernova shock waves, that $\eta_{\rm p}$ might be as large as $\sim 10\%$. Since particles are accelerated in the considered scenario at the shock, it is assumed that they obtain a simple power law spectrum with the spectral index is usually considered in the range $2-3$. 
We realise that nonlinear effects may be important in diffusive shock acceleration causing the hardening of the spectrum with energy (e.g. Baring et al.~1999). However, we are interested in the basic
radiation features of particles inside the open clusters. The impression on the results obtained with the assumption on the more complicated spectra can be reached by comparing the results obtained with simple power law spectra and some range of spectral indexes.
The limits on the maximum energies of particles in terms of the adopted here acceleration scenario has been considered many times in the past by comparing the energy losses of particles with their acceleration efficiency. In the case of electrons we refer to the recent paper by Bednarek \& Sitarek~(2007) in which their acceleration and losses in specific
situation of the globular cluster are considered. Here we only repeat the final results and
extend them for the case of interaction of electrons with the matter (which was not the case of the globular clusters).

By comparing the acceleration time scale (see e.g. Eq.~7 in Bednarek \& Siterek~2007) with the synchrotron energy loss time scale, we get the limit on the energy of accelerated electrons due to the synchrotron losses,
\begin{eqnarray}
E_{\rm syn}^{\rm max}\approx 58(\xi_{-5}/B_{-5})^{1/2}~~~{\rm TeV.} 
\label{eq3}
\end{eqnarray}
where  $B = 10^{-5}B_{-5}$ G is the magnetic field strength at the acceleration region, $R_{\rm L}$ is the Larmor radius of particles,  $E=1E_{\rm TeV}$ TeV is its energy, $c$ is the velocity of light, and $\xi = 10^{-5}\xi_{-5}$ is the so called acceleration coefficient ($\xi\le 1$). 

Important question concerns the value of the acceleration efficiency. Let us consider, as 
an example, the open cluster Berk 87. The diffuse hard X-ray emission has been detected from its direction below a few keV by EXOSAT (Warwick et al.~1988). If this X-ray emission is produced by electrons in the synchrotron process, then the characteristic energies of synchrotron photons can be estimated from,

\begin{eqnarray}
\varepsilon\approx m_{\rm e}c^2\gamma^2(B/B_{\rm cr})\approx 4.6\times 10^{-4}B_{-5}E_{\rm TeV}^2~~~{\rm keV},
\label{eq4}
\end{eqnarray}

\noindent
where $B_{\rm cr} = 4.4\times 10^{13}$ G, and $m_{\rm e}c^2$ is the electron rest mass.
By introducing Eq.~\ref{eq3} into Eq.~\ref{eq4}, we estimate the maximum energies of synchrotron photons, which can be produced in Berk 87, on $\varepsilon_{\rm max}\sim 1.5\xi_{-5}$ keV. 
Therefore, for $\varepsilon\sim 5$ keV, the acceleration coefficient has to be of the order of
$\sim 3\times 10^{-5}$. On the other hand, the theory of acceleration of particles at non-relativistic shocks gives the estimate of the acceleration coefficient $\xi\approx 0.1 (v/c)^2$, where $v$ is the velocity of the plasma through the shock. The wind velocity measured from the  WR star, WR 142 (present in Berk 87), is $\sim$5200 km s$^{-1}$. Then, we obtain the value $\xi\sim 3\times 10^{-5}$, consistent with the above estimate. Therefore, we conclude that synchrotron emission from electrons accelerated at the shock in the WR star wind can be responsible for the observed diffuse X-ray emission from Berk 87 at energies up to a few keV. 
The energy losses of electrons on the inverse Compton scattering (ICS) of thermal photons produced by the stars has been also considered 
in Bednarek \& Sitarek~(2007). However, they are not able to limit the acceleration of electrons  (see Fig.~1 on which we compare the times scales for different processes of leptons). However, IC scattering can be important during the cooling phase of electrons inside the open cluster.

Since large amount of distributed matter is expected inside the open clusters, we should also consider the energy losses of electrons on the bremsstrahlung process. In the relativistic case, they can be approximated by (Lang~1999),
\begin{eqnarray}
\left({{dE}\over{dt}}\right)_{\rm br}\approx  m_{\rm p}c^3NE/X\approx 
7.8N_4E_{\rm TeV}~~~{\rm eV~s^{-1}},
\label{eq6}
\end{eqnarray}
\noindent
where, $X$ is the radiation length equal to $X = 62.8$ gram cm$^{-2}$ for hydrogen atoms,
$m_{\rm p}$ is the proton mass, and $N=10^4N_4$ cm$^{-3}$ is the number density of target protons. By comparing acceleration efficiency of electron's with their bremsstrahlung energy losses, we find that they are usually not able to determine the maximum energies of electrons (see Fig.~1).

Leptons injected inside the open cluster diffuse gradually out of it. Let us assume that the diffusion process can be approximated by the the Bohm diffusion. Such assumption migth be justified in the case of very turbulent media also characteristic to the open clusters is which multiple stellar winds are present and supernova explosions occured. 
The characteristic time spent by leptons with energy, E, inside the open cluster with characteristic radius $R_{\rm oc}$, can be estimated by,   
\begin{eqnarray}
t_{\rm diff} = R_{\rm oc}^2/D_{\rm diff}\approx  3\times 10^{12}R_{\rm oc}^2B_{-5}/E_{\rm TeV}~~~{\rm s},
\label{eq10}
\end{eqnarray}
\noindent
where $D_{\rm diff} = R_{\rm L}c/3$ is the diffusion coefficient, $R_{\rm L} =p/eB_{\rm GC}\approx 3\times 10^{14} E_{\rm TeV}/B_{-5}$ cm is the Larmor radius of electrons in the magnetic field of the open cluster, and $R_{\rm oc}$ is the core radius of the open cluster. Note that diffusion time of electrons with energies 
$< 1$ TeV is longer than the characteristic age of the WR star, $t_{\rm WR}\approx 3\times 10^5$ yrs, responsible for the acceleration of particles. So then, leptons radiate inside the open cluster since these ones with larger energies cool much faster on radiation processes.
The age of the open clusters, e.g. Berk 87 ($\sim 2\times 10^6$ yrs), is at least an order of magnitude larger than the age of the WR star.

We conclude that electrons are accelerated at the shock wave in the WR stellar wind to maximum energies limited by the synchrotron process. However, the production of $\gamma$-rays in the IC and bremsstrahlung process is important for electrons with lower energies.

\begin{figure}
\vskip 6.truecm
\includegraphics{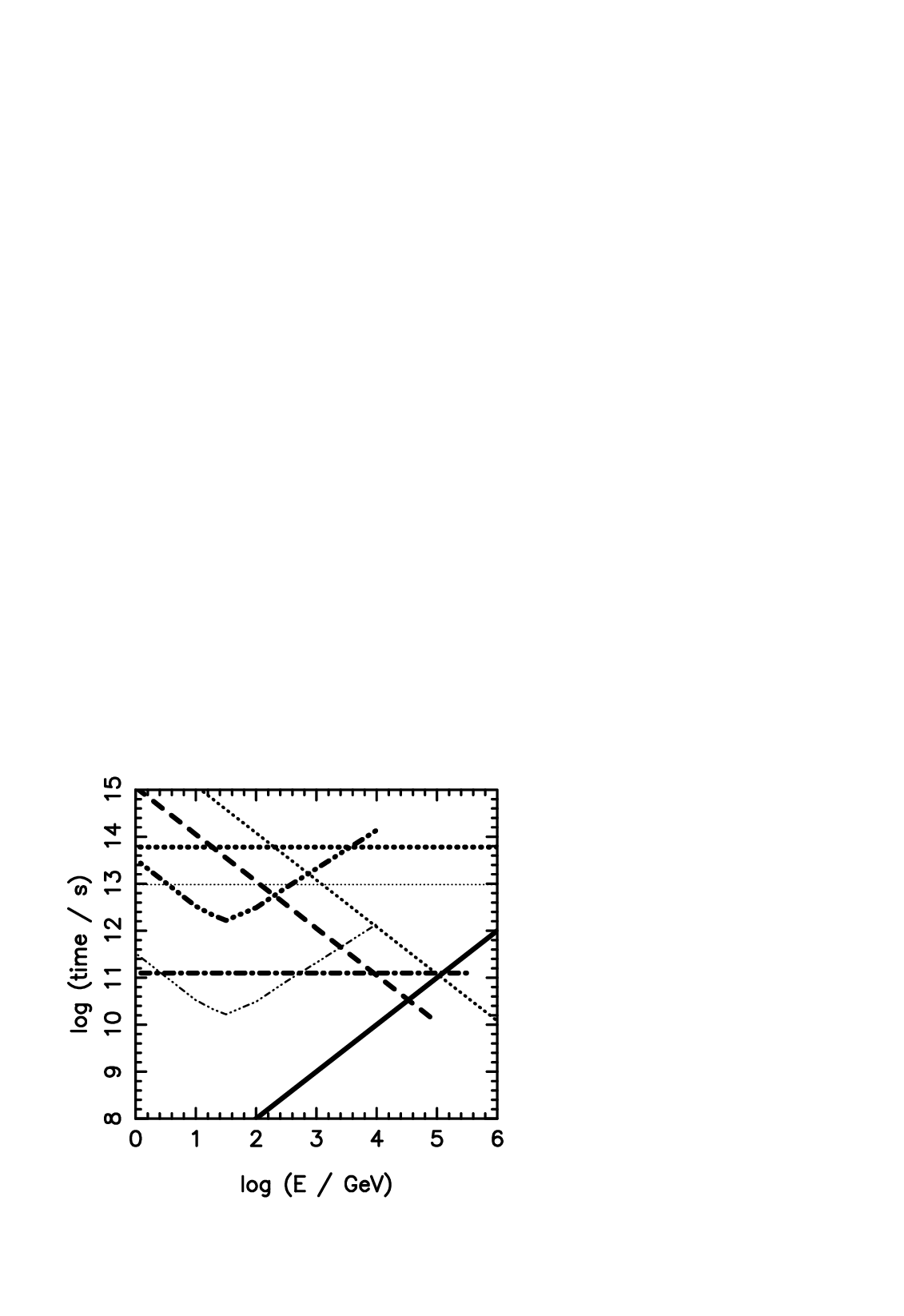}
\caption{The characteristic time scales for electrons on different radiation processes, 
synchrotron (dashed line), bremsstrahlung  (dot-dashed), ICS on stellar radiation (thick 
triple-dot-dashed), and for their energy gains from acceleration 
mechanism (thick solid) are compared with the diffusion time scale of electrons from 
the open cluster calculated for $R_{\rm oc} = 2$ pc (middle dotted), the age of the WR star $3\times 10^5$ yrs (thin dotted), and the age of 
the open cluster $2\times 10^6$ yrs (thick dotted). The parameters which determine the energy losses and gains are the following: $B = 10^{-5}$ G, $N = 10^4$ cm$^{-3}$, $\xi = 10^{-5}$, and $\mu = 1$ (thick triple-dot dashed curve) and 100 (thin triple-dot dashed).
}
\label{fig1}
\end{figure}

Also hadrons can be accelerated at the shock waves in the massive star winds. They mainly lose energy in the interaction with the matter inside the open cluster.  The approximate energy losses on pion production in proton-proton (p-p) collisions (in the relativistic case $E_{\rm p}\gg m_{\rm p}c^2$) can be estimated from,
\begin{eqnarray}
\left({{dE}\over{dt}}\right)_{\rm pp}\approx  c\sigma_{\rm pp}\kappa_\pi NE_{\rm p}\approx 
4.5N_4E_{\rm TeV}~~~{\rm eV~s^{-1}},
\label{eq11}
\end{eqnarray}
\noindent
where $E_{\rm p} = 1E_{\rm TeV}$ TeV is the proton energy, $\sigma_{\rm pp}\approx 3\times 10^{-26}$ cm$^{-2}$ and $\kappa_\pi \approx 0.5$ are the cross section and the in-elasticity coefficient for p-p interaction. By comparing the acceleration energy gains (Bednarek \& Sitarek~2007) with the energy loss time scale for protons (Eq.~\ref{eq11}), we get the absolute upper limit on the energies of accelerated protons, 
\begin{eqnarray}
E_{\rm pp}^{\rm max}\approx 670B_{-5}/N_4~~~{\rm TeV.} 
\label{eq12}
\end{eqnarray}
\noindent
However, the maximum energies of hadrons can be also limited by the characteristic time scales resulting from the confinement of hadrons inside the open cluster due to their diffusion process (this also depends on the energy of hadron), and the age of the acceleration mechanism defined  by the age of the WR star ($\sim 3\times 10^5$ yrs). Note, that it this picture the dimension of the open cluster is comparable to the dimensions of the shocks.
Let us at first consider the diffusion of relativistic protons through the open cluster.
The diffusion time scale of protons (in the Bohm approximation) is described as in the case of electrons by Eq.~\ref{eq10}. By comparing it with the energy gain time scale, $\tau_{\rm acc} = E_{\rm p}/(dE/dt)_{\rm acc}$, we get another limit on the maximum proton energies,
\begin{eqnarray}
E_{\rm dif}^{\rm max}\approx 32R_{\rm oc}B_{-5}\xi_{-5}^{1/2}~~~{\rm TeV.} 
\label{eq13}
\end{eqnarray}
\noindent
where, $R_{\rm oc}$ is the cluster radius in parsecs. This limit is more stringent than the limit obtained based on the energy losses of protons in hadronic collisions. {Note, moreover that the Larmor
radii of particles with such energies are much smaller than the dimension of the cluster.}

In order to check whether the process of energy transfer from protons to radiation is efficient, we  estimate the collision rate of protons with the matter inside the open cluster, 
\begin{eqnarray}
N_{\rm col}^{\rm pp} = c t_{\rm diff} \sigma_{\rm pp} N\approx 110N_4B_{-5}/E_{\rm TeV}, 
\label{eq15}
\end{eqnarray}
\noindent
where $t_{\rm diff}$ is given by Eq.~\ref{eq10}, and the radius of the open cluster is taken equal to $R_{\rm oc} = 2$ pc (as in Berk 87). It is clear that protons with large energies are not able to lose very efficiently energy in the interactions with the matter during their propagation through the open cluster provided that $E_{\rm TeV} > 10^2 N_4B_{-5}$. Protons with large enough energies escape from the open cluster with only partial energy losses. However, they can be captured inside the high density clouds. Therefore, we should observe a break in the $\gamma$-ray spectrum expected from the open cluster at energies corresponding to energies of hadrons mentioned above.
On the other hand, hadrons with energies above these values, should produce efficiently $\gamma$-rays in their interactions with the matter of the dense clouds surrounding the open cluster.
The break in the $\gamma$-ray spectrum observed from the open cluster may in fact appear at larger energies than estimated above if their escape from the open cluster is more efficient, e.g. due to the partially ordered magnetic field. 

\begin{figure}
\vskip 6.truecm
\includegraphics{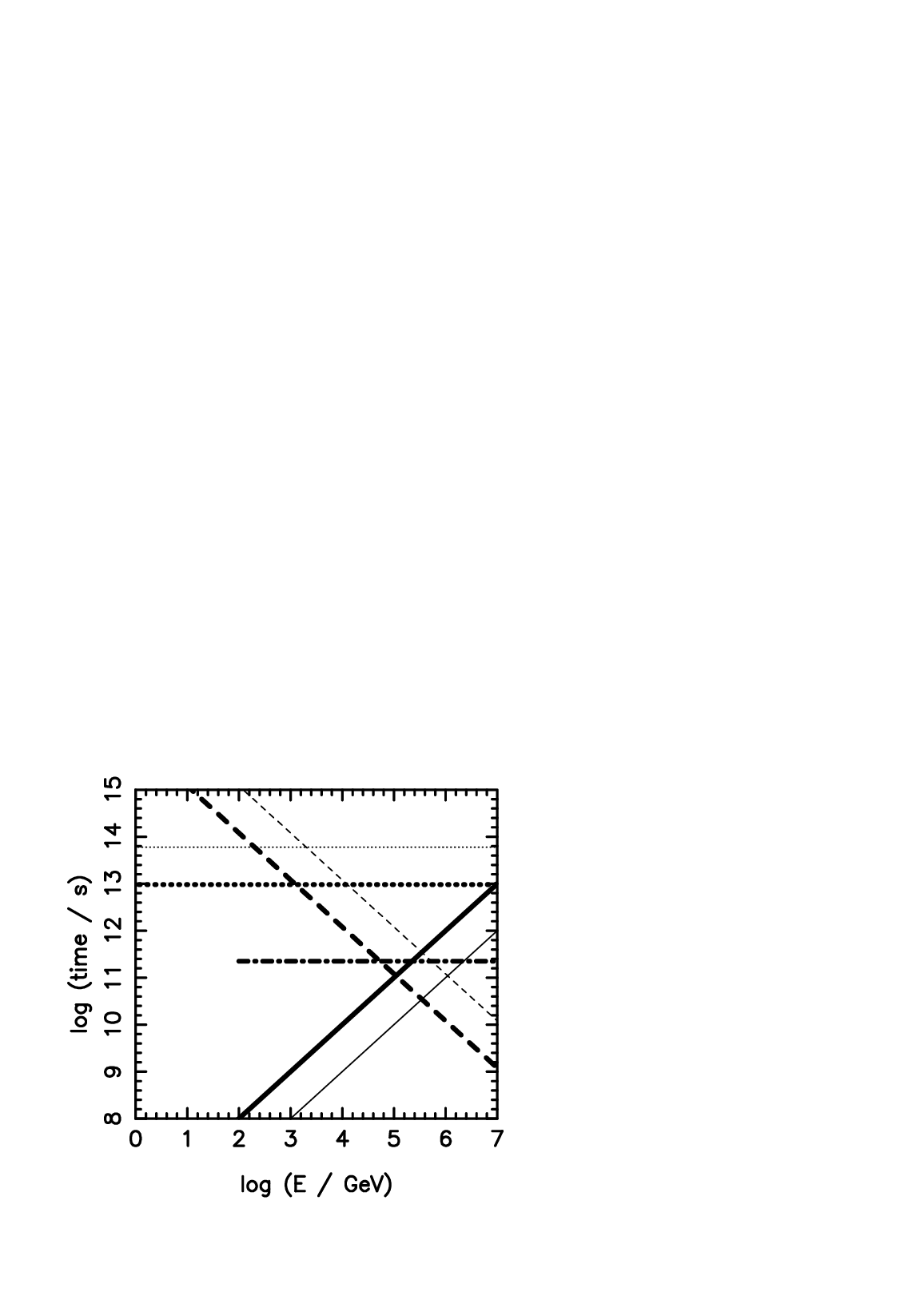}
\caption{As in Fig.~\ref{fig1} but for relativistic protons injected into the open cluster. 
The time scale for protons losing energy on pion production in collisions with the matter 
(dot-dashed line). Their time scales for energy gains from the acceleration mechanism are 
shown for the magnetic field strength $B = 10^{-5}$ G (thick solid line) and $B = 10^{-4}$ G 
(thin solid line). The diffusion time 
scale of hadrons through the open cluster is shown by the dashed line, the age of the WR star, 
and the age of the open cluster are marked by the thick and thin dotted lines, respectively. The other parameters describing the open cluster are this same as applied in Fig.~\ref{fig1}.}
\label{fig3}
\end{figure}

The maximum energies of hadrons are determined by their 
diffusion outside the open cluster. For the average magnetic field strength inside the open 
cluster equal to $B = 10^{-5}$ G, hadrons can be accelerated to energies above 
$E_{\rm max}^{\rm p}\sim 10^5$ GeV.

\section{Gamma-ray production inside open clusters}

\begin{figure*}
\vskip 14.truecm 
\includegraphics{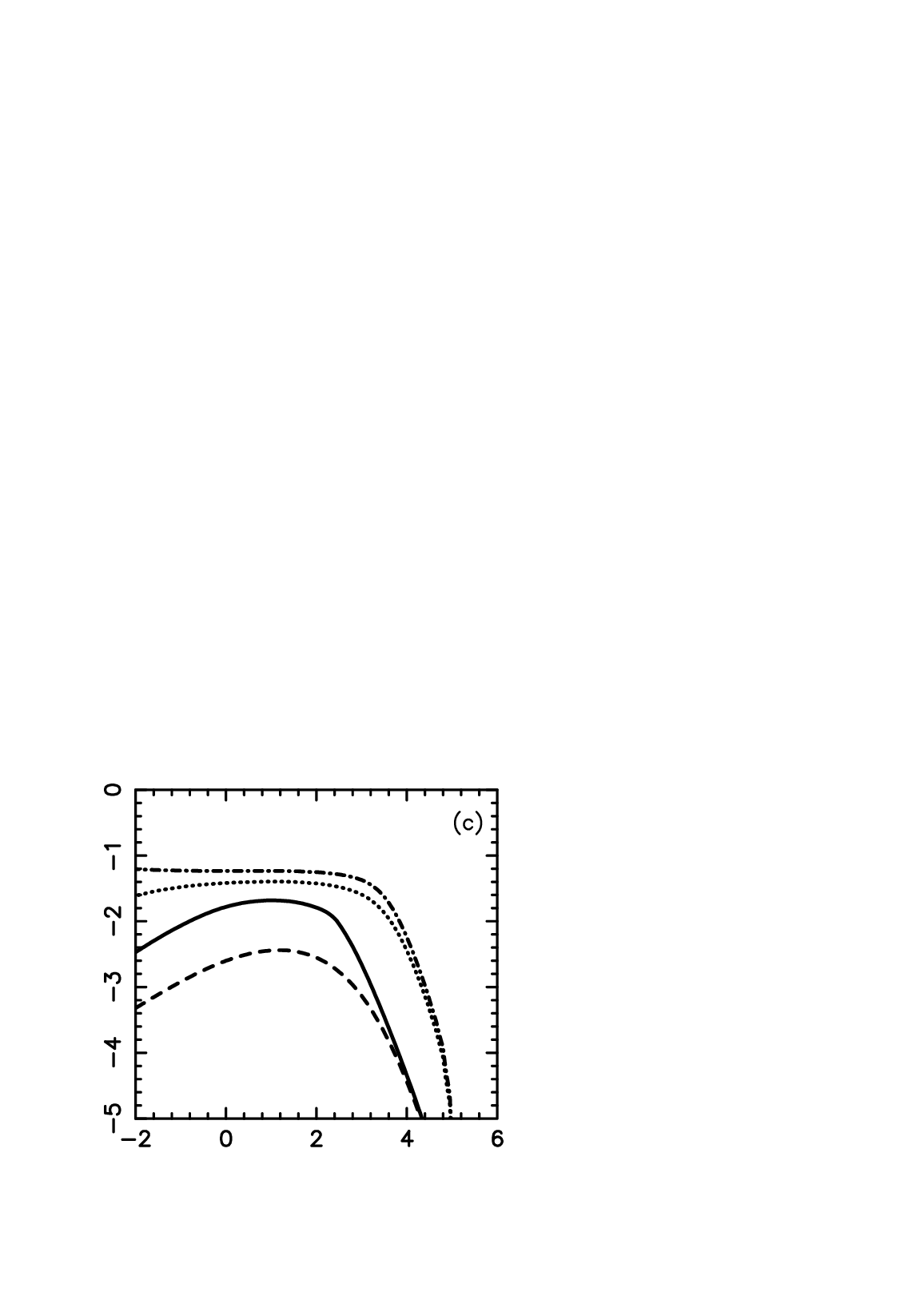} 
\includegraphics{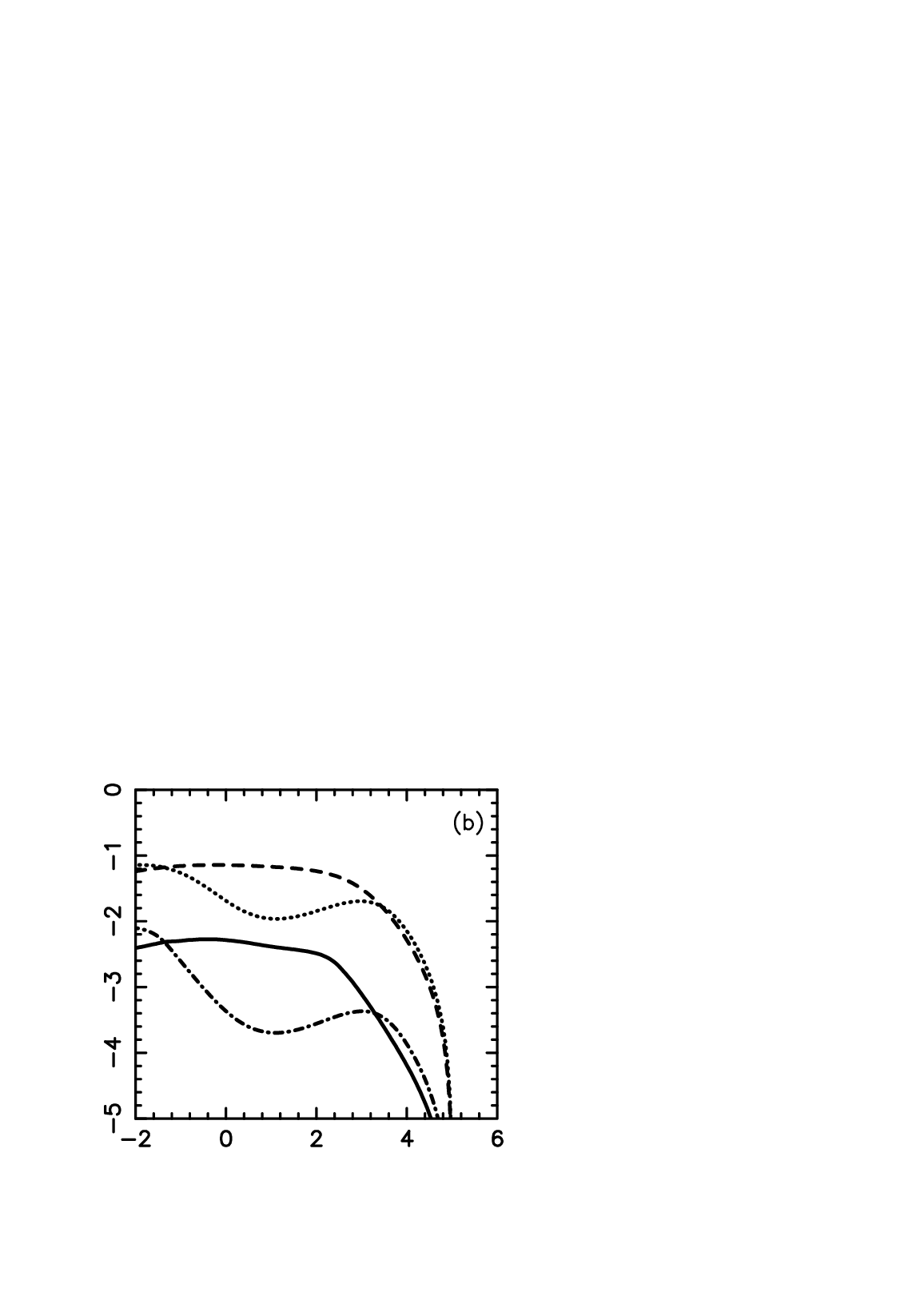}
\includegraphics{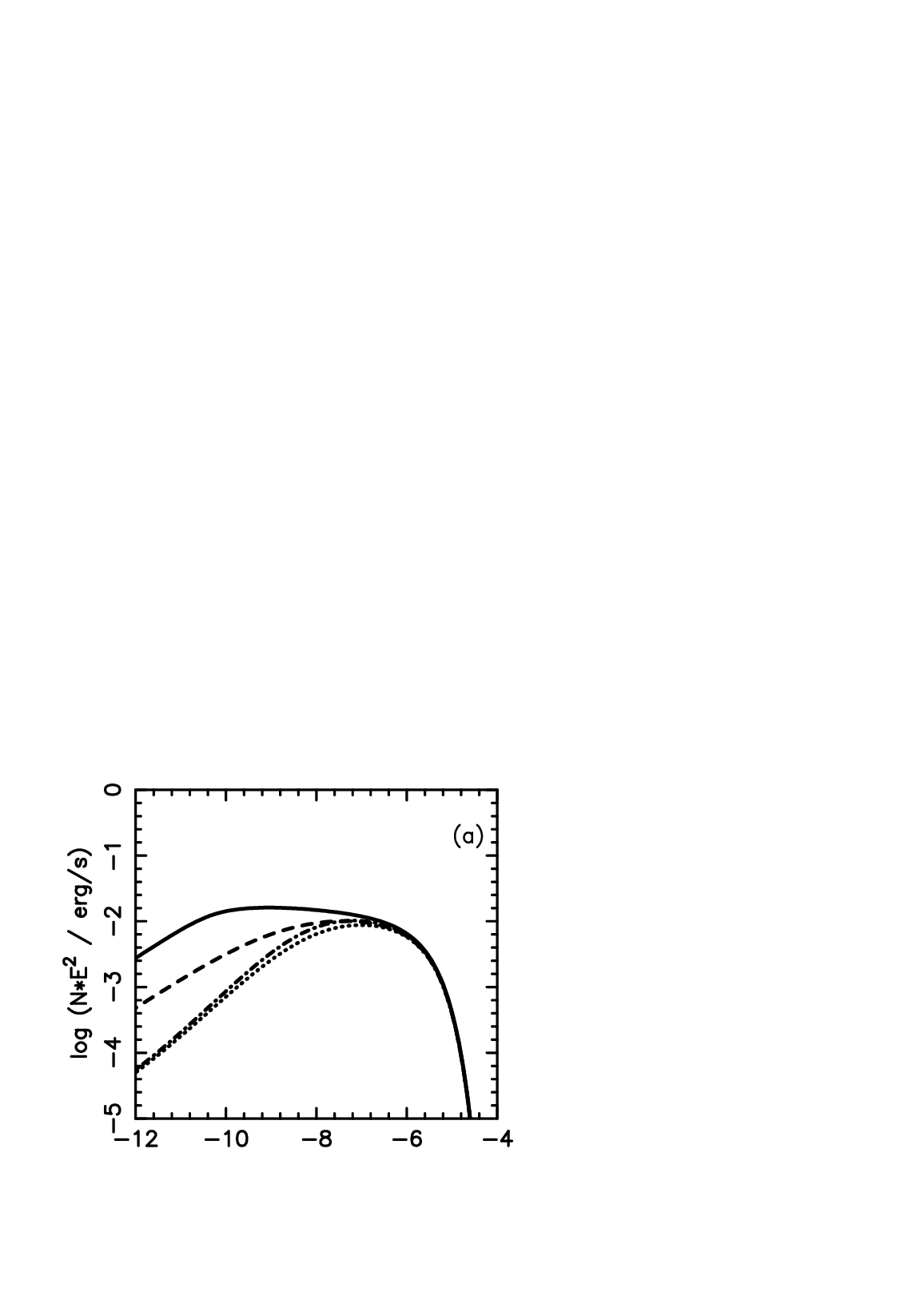} 
\includegraphics{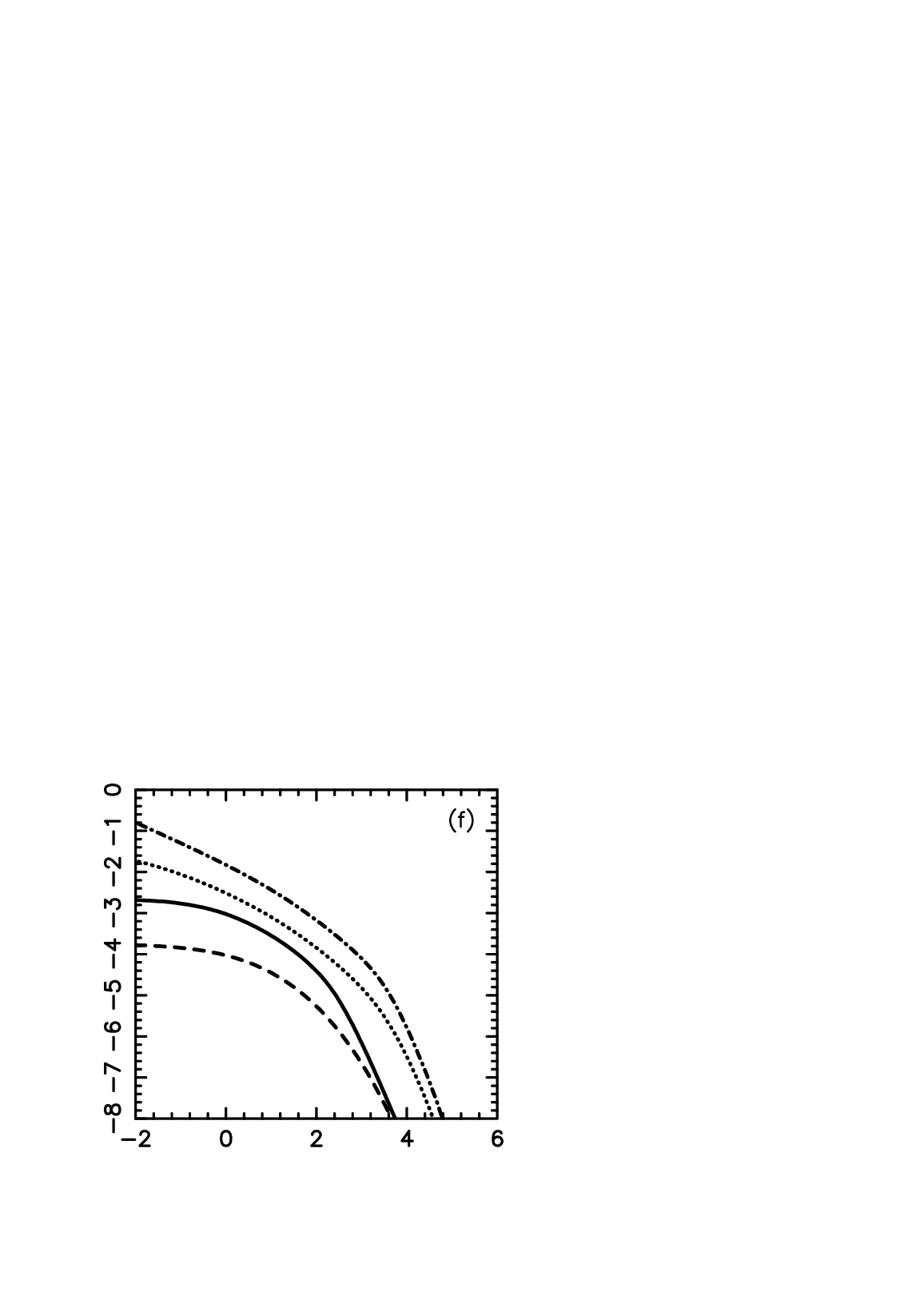} 
\includegraphics{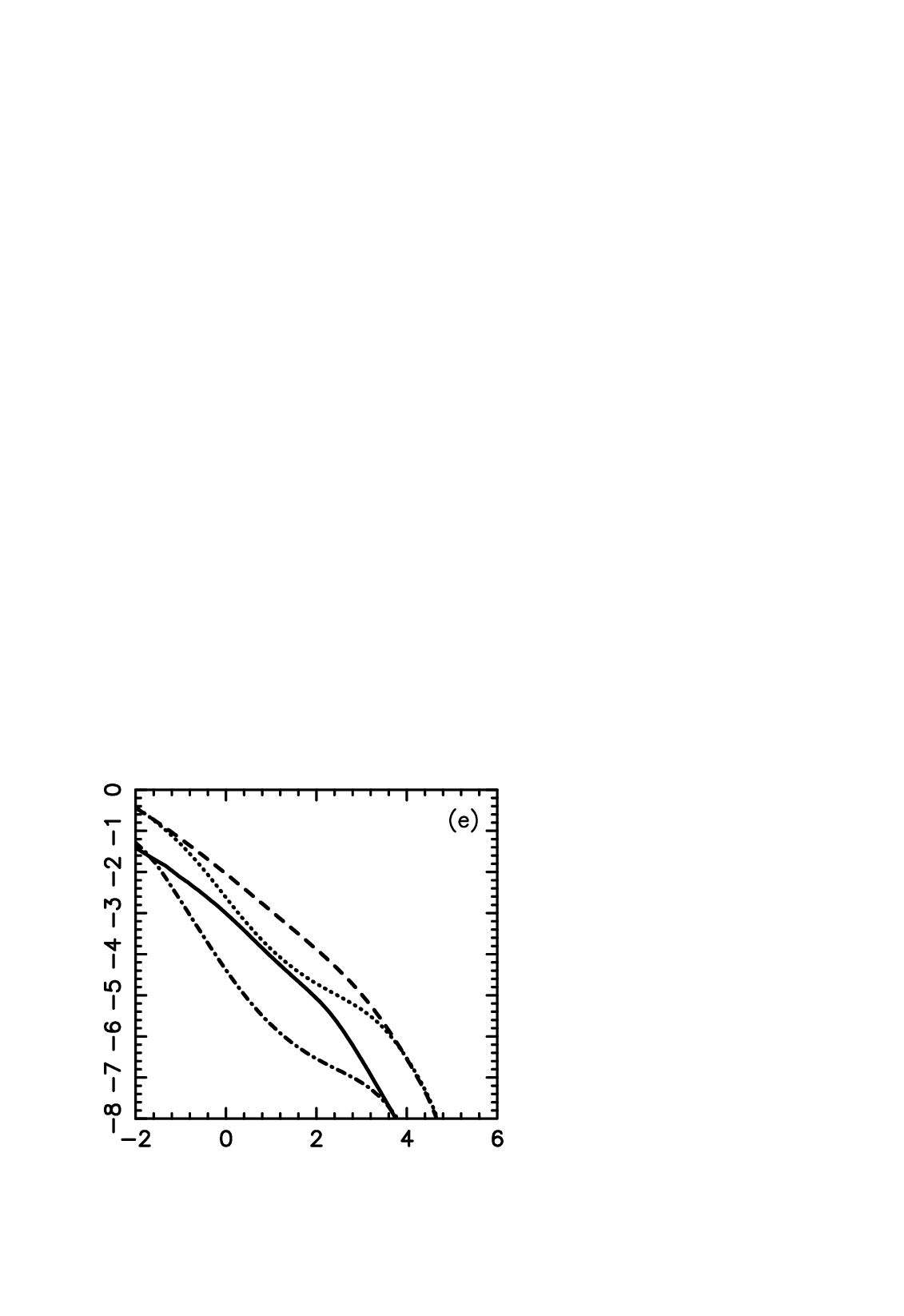} 
\includegraphics{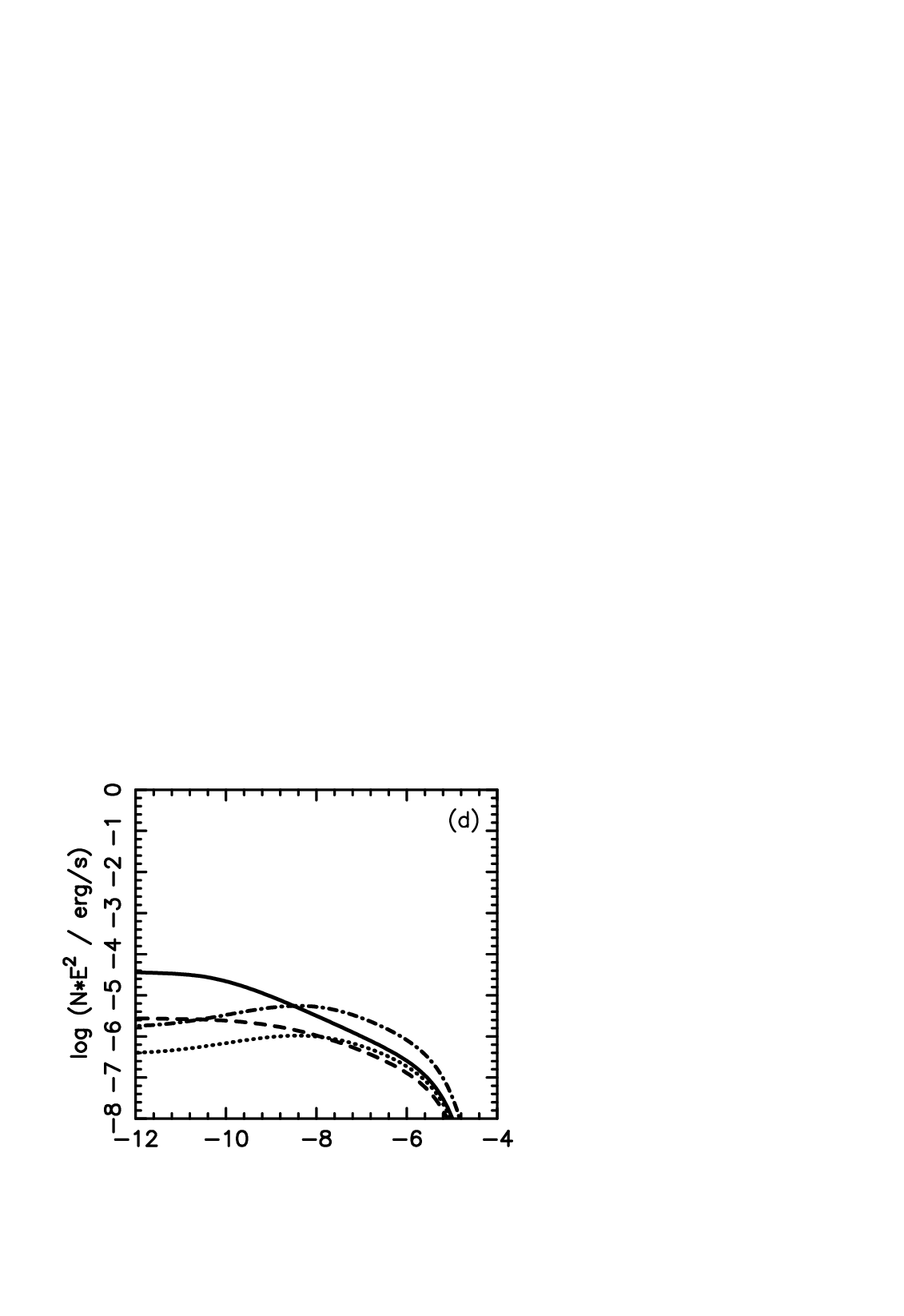} 
\includegraphics{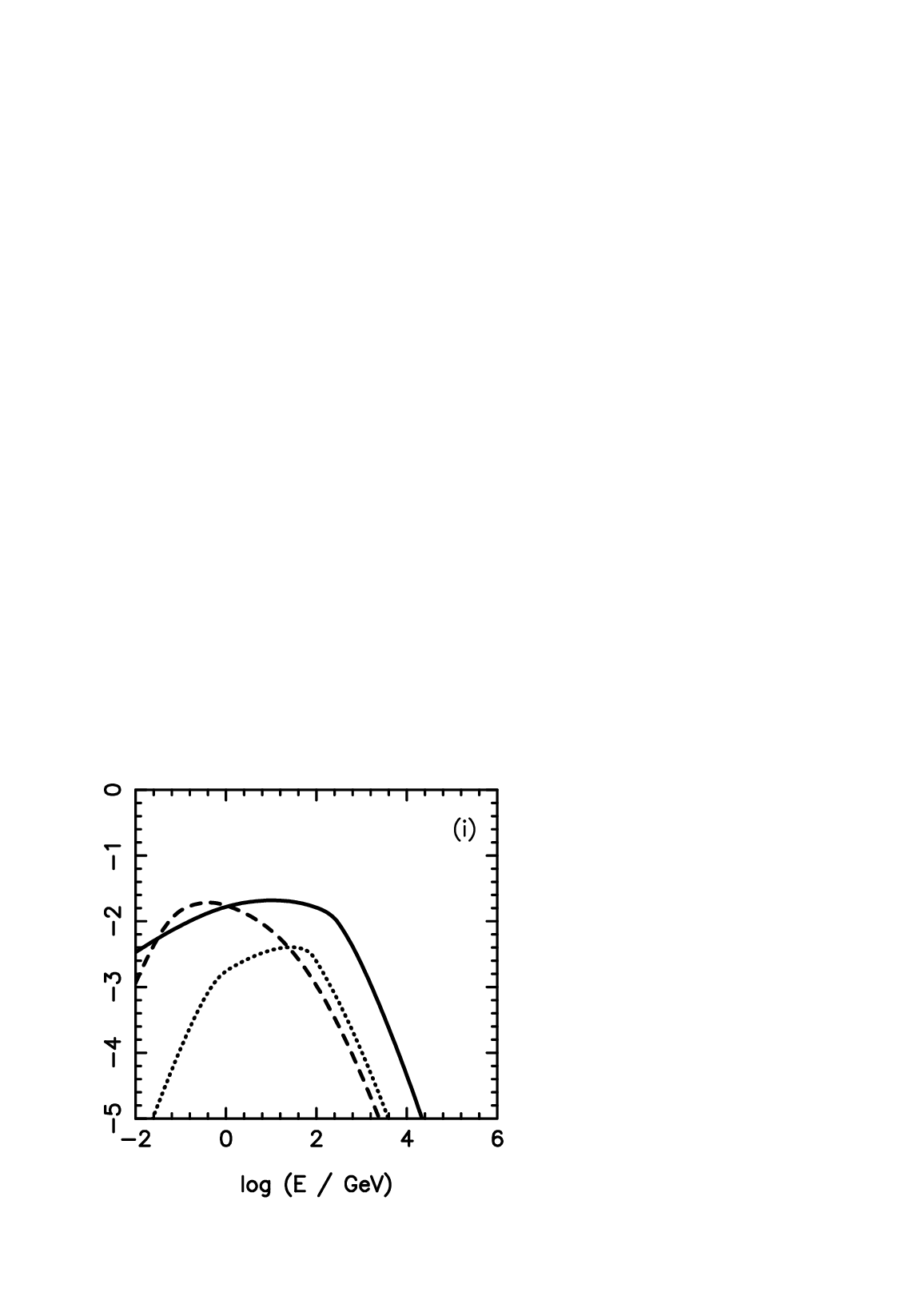} 
\includegraphics{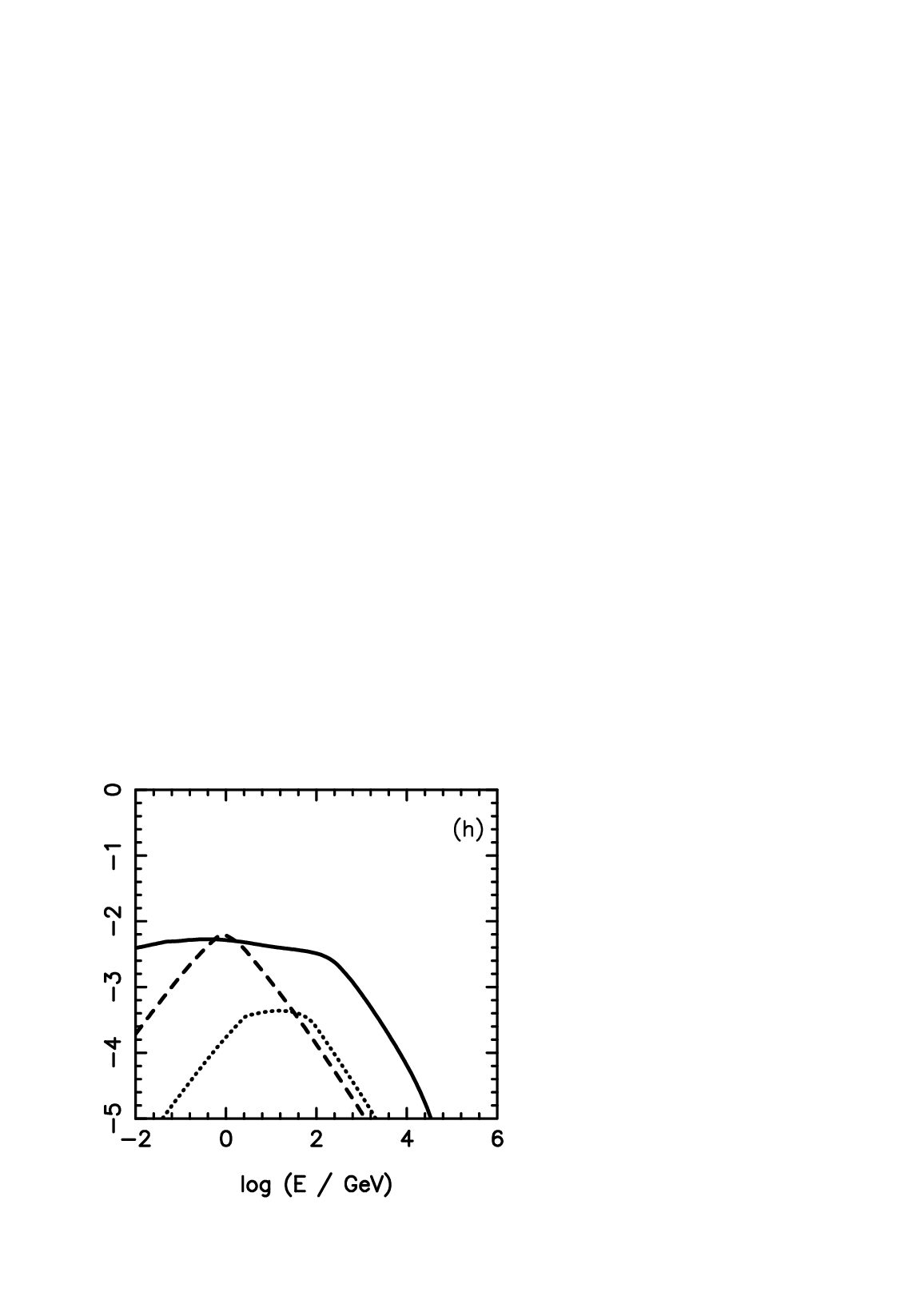} 
\includegraphics{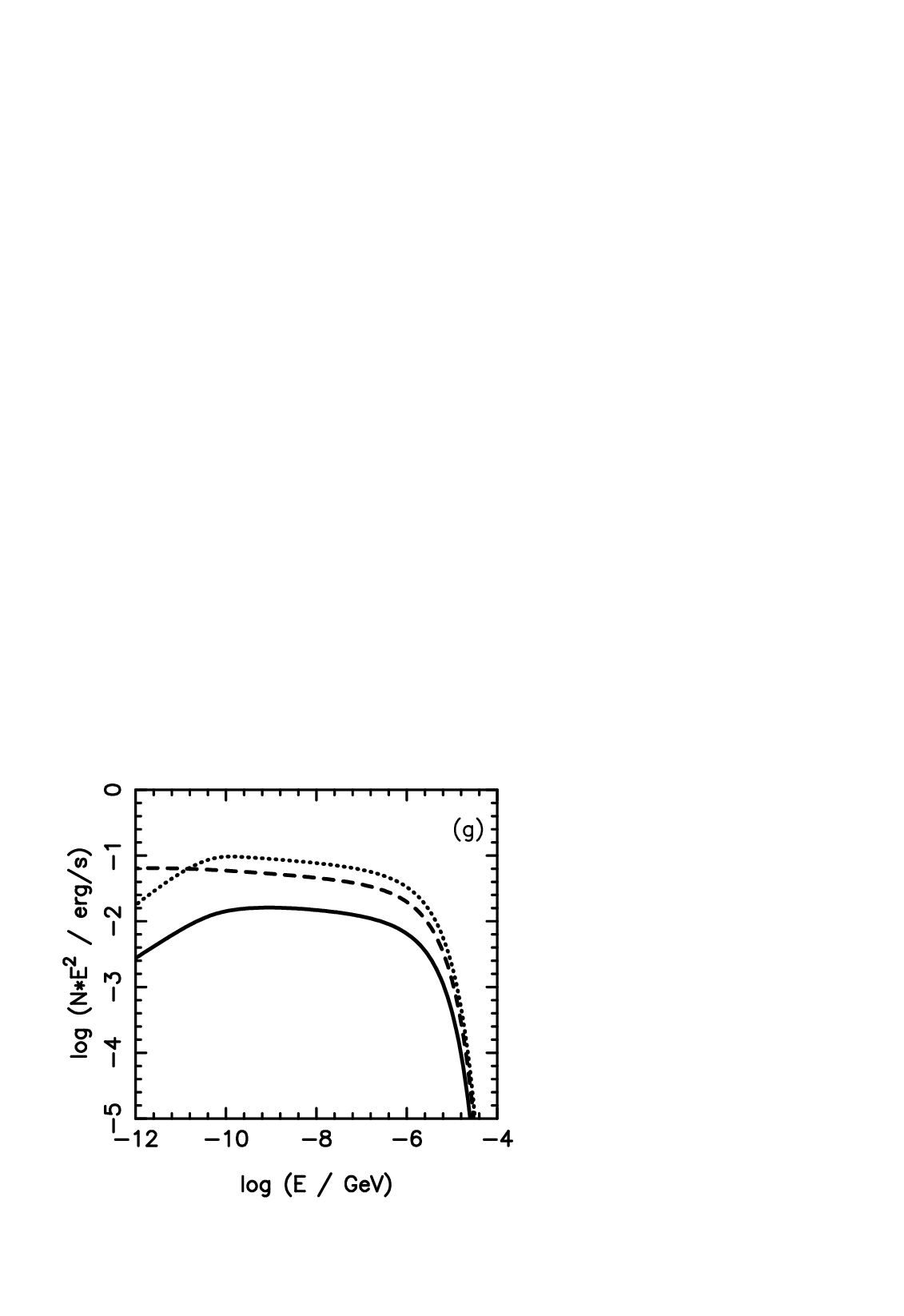} 
\caption{The differential photon spectra (multiplied by the energy squared, SED) 
produced in the synchrotron ((a), (d) and (g)), bremsstrahlung ((b), (e) and (h)), and IC processes ((c), (f) and (i)) are calculated for the power law spectra of injected electrons with 
the spectral index 2.1 (upper panel) and 3 (middle panel) for the low energy cut-off in the 
electron spectrum at $E_{\rm min} = 0.01$ GeV and the high energy cut-off at $E_{\rm max}$ (obtained for $B = 10^{-5}$ G, see Eq.~\ref{eq3}), and different densities of matter ($\rho$) and  stellar radiation field (defined by $\mu$) inside the open cluster: $\mu = 1$ and $\rho = 10$ cm$^{-3}$ (solid curves), $\mu = 1$ and $\rho = 10^3$ cm$^{-3}$ (dashed curves), $\mu = 100$ and $\rho = 10$ cm$^{-3}$ (dot-dashed curves), $\mu = 100$ and $\rho = 10^3$ cm$^{-3}$ (dotted curves). The spectra produced in these radiation processes are also shown for different low and high energy cut-offs and fixed values of $\mu = 1$ and 
$\rho = 10$ cm$^{-3}$: $E_{\rm min} = 0.01$ GeV and $E_{\rm max}$ obtained for $B = 10^{-5}$ G 
(solid curves); $E_{\rm min} = 1$ GeV and  $E_{\rm max}$ obtained for $B = 10^{-4}$ G 
(dashed curves); $E_{\rm min} = 10$ GeV and $E_{\rm max}$ obtained for $B = 10^{-5}$ G 
(dotted curves). The other parameters of the open cluster are taken as derived for Berk 87
(see section 2.1 for details).} 
\label{fig2}
\end{figure*}

Based on the above analysis of the radiation and propagation processes we conclude that electrons 
should lose most of their energy already inside the low density cavity created by the massive stars.
We assume that its size is comparable to the core radius of the open cluster. However, the most energetic hadrons at first, diffuse outside this low density cavity and next, interact 
with dense matter surrounding the open cluster. Therefore, it is expected that $\gamma$-rays 
produced by electrons should arrive directly from the open cluster but the most energetic 
$\gamma$-rays produced in hadronic collisions should come from a more extended region surrounding the open cluster, i.e. from the giant molecular clouds. In the present calculations electrons and hadrons are 
injected uniformly during the lifetime of the Wolf-Rayet star (which is of the order of $\sim 
3\times 10^5$ yrs). In fact, massive WR stars might also appear in the past inside 
the open cluster. Therefore, especially  lower energy $\gamma$-ray emission observed from the open cluster may represent a combination of the emission from the WR stars visible at present and from some WR stars which have been already exploded as a supernovae.
However, the contribution of these 'old' WR stars is very difficult to take into account
due to the lack of any observables which might allow us to fix their number and parameters.
Below, we only show the results of calculations of high energy radiation produced by electrons and hadrons from the presently observed massive stars in specific open clusters.

All three important radiation processes have to be taken into account 
when calculating the $\gamma$-ray spectra produced by electrons inside a specific open cluster. 
Diffusion time scale of electrons through the open cluster is 
typically much longer than the radiation time scales of electrons (Fig. \ref{fig1}). Therefore, electrons accelerated at the shock in the WR star wind lose most of their energy inside the open cluster. We calculate the differential photon spectra produced by electrons for different parameters of the open cluster and the 
parameters describing the injection spectrum of electrons. The spectra, shown in Fig.~\ref{fig2}, are calculated for the power law injection spectrum of electrons with the spectral indices $2.1$ (upper figures) and $3$ (middle figures), the low energy cut-off at $10^{-2}$ GeV and the high energy cut-off determined by the synchrotron energy losses in the magnetic field. The spectra are normalized to 1 erg s$^{-1}$. 
Specific curves in these figures are obtained for different effective radiation field inside 
the open cluster produced by the massive stars and the average density of the matter inside 
the volume of the open cluster. We investigate the cases in which density of the radiation 
field and matter can change by a factor of one hundred around the basic parameters $\mu = 1$ 
and $\rho = 10^2$ cm$^{-3}$.

For electrons injected with flat spectra (spectral index $\alpha = 2.1$) and the parameters of the medium from the lower part of the considered range, the $\gamma$-ray spectrum
produced in the IC process dominates over the $\gamma$-ray spectrum produced in the bremsstrahlung process and the lower energy photon spectrum (from radio to X-rays) produced in the synchrotron process (e.g. see solid curves in Fig.~\ref{fig2}abc). 
However, electrons with different energies contribute at different level to specific parts of 
the photon spectrum. The highest energy electrons radiate
mainly on synchrotron process and electrons with energies between a few GeV and a few hundred GeV lose energy mainly in IC process. However, relative contribution of specific radiation processes can change drastically with the change of the basic parameters of the open cluster. If the effective density of stellar 
radiation field is larger (see the cases marked by the dot-dashed and dotted curves), 
the $\gamma$-ray spectrum in the broad energy range (up to a few TeV) is well described by a simple power law with the spectral index close to 2 (as expected from the complete cooling case 
of electrons in the T and IC regimes for electrons injected with the spectral index close to 2). 
The $\gamma$-ray spectra produced for these parameters in the bremsstrahlung process are 
on a much lower level since assumed density of matter inside the open cluster is lower. 
The relative contribution of specific radiation processes can change drastically for electrons 
injected with much steeper spectrum, e.g. for $\alpha = 3$ (compare the upper and middle 
panels in Fig.~\ref{fig2}). For such steep electron spectra, synchrotron emission (from radio up to X-rays) drops by a few orders of magnitudes in respect to the IC and bremsstrahlung emission (in the GeV-TeV energy range).

Dependence of the photon spectra on the low and high energy cut-offs in the injected spectrum of electrons is investigated in the bottom figures (see Fig.~\ref{fig2}). 
Here we consider the open cluster with the stronger magnetic field 
(i.e. $B = 10^{-4}$ G) which, as evident from Eq.~\ref{eq3}, should allow acceleration of electrons to lower maximum energies). Note that the cut-off in the electron spectrum from the lower energies is not well determined. It is difficult to motivate its presence theoretically on the present state on knowledge. 
However, we consider such general spectra since they seems to give better description of the flat
$\gamma$-ray spectrum observed at GeV energies from the open clusters.
The electron spectra with the low energy cut-off at $E_{\rm min} = 1$ and 10 GeV are considered (see dashed and dotted curves in Fig~\ref{fig3}, respectively). As an example, we 
assume that electrons have the spectral index $\alpha = 2.1$ and the medium of the open cluster is characterized by $\mu = 1$ and $\rho = 10$ cm$^{-3}$. For such spectral index, the energy is almost equally distributed throughout the electron spectrum. Therefore, the synchrotron spectra, produced by the largest energy electrons, clearly dominate over the $\gamma$-ray spectra produced in the IC and bremsstrahlung processes since electrons can not cool completely during the age of the WR star. Note, 
that the $\gamma$-ray spectra produced for these parameters are rather steep even for the flat 
spectrum of injected electrons due to the dominance of the synchrotron cooling. 

\begin{figure}
\vskip 13.5truecm
\includegraphics{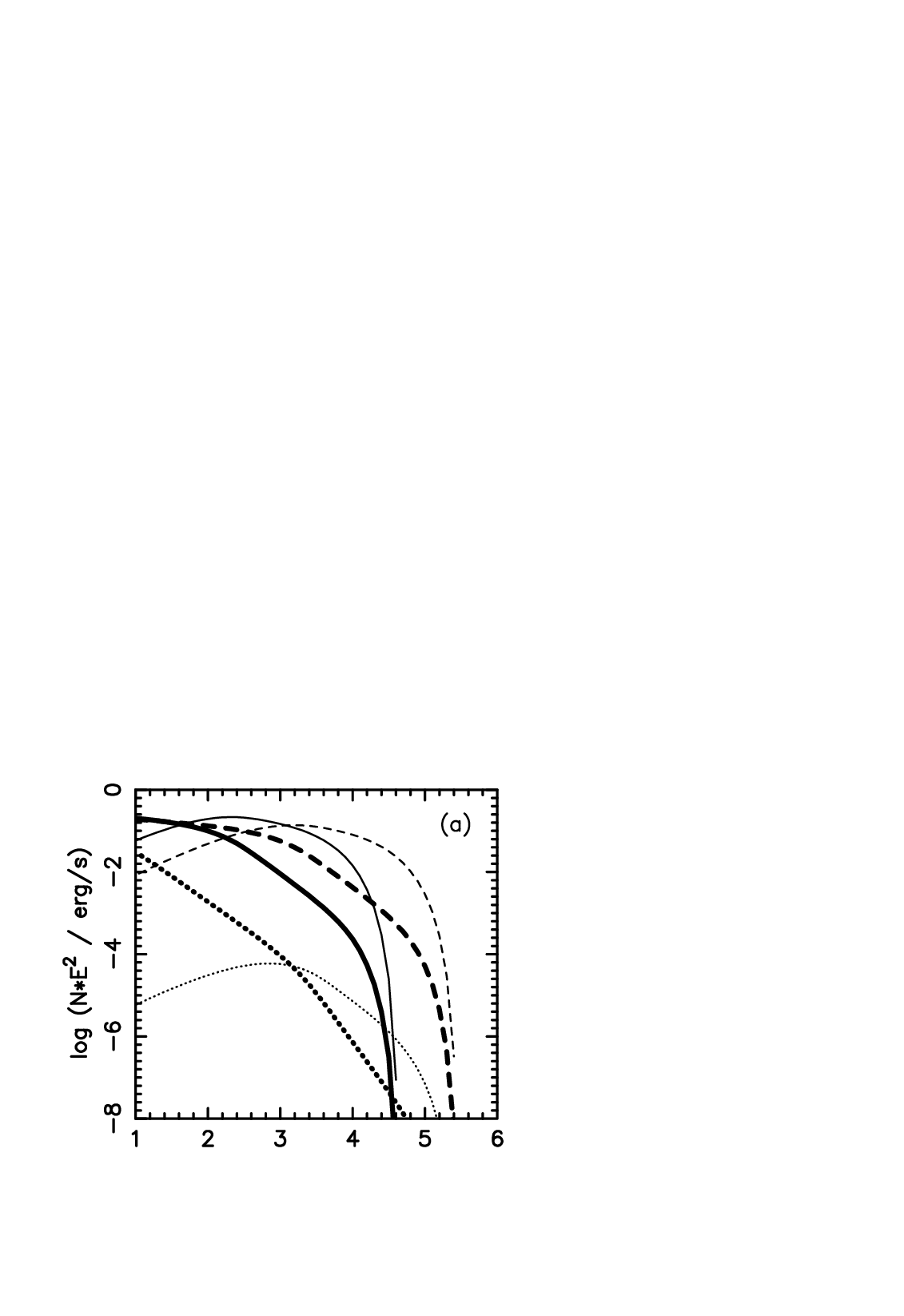}
\includegraphics{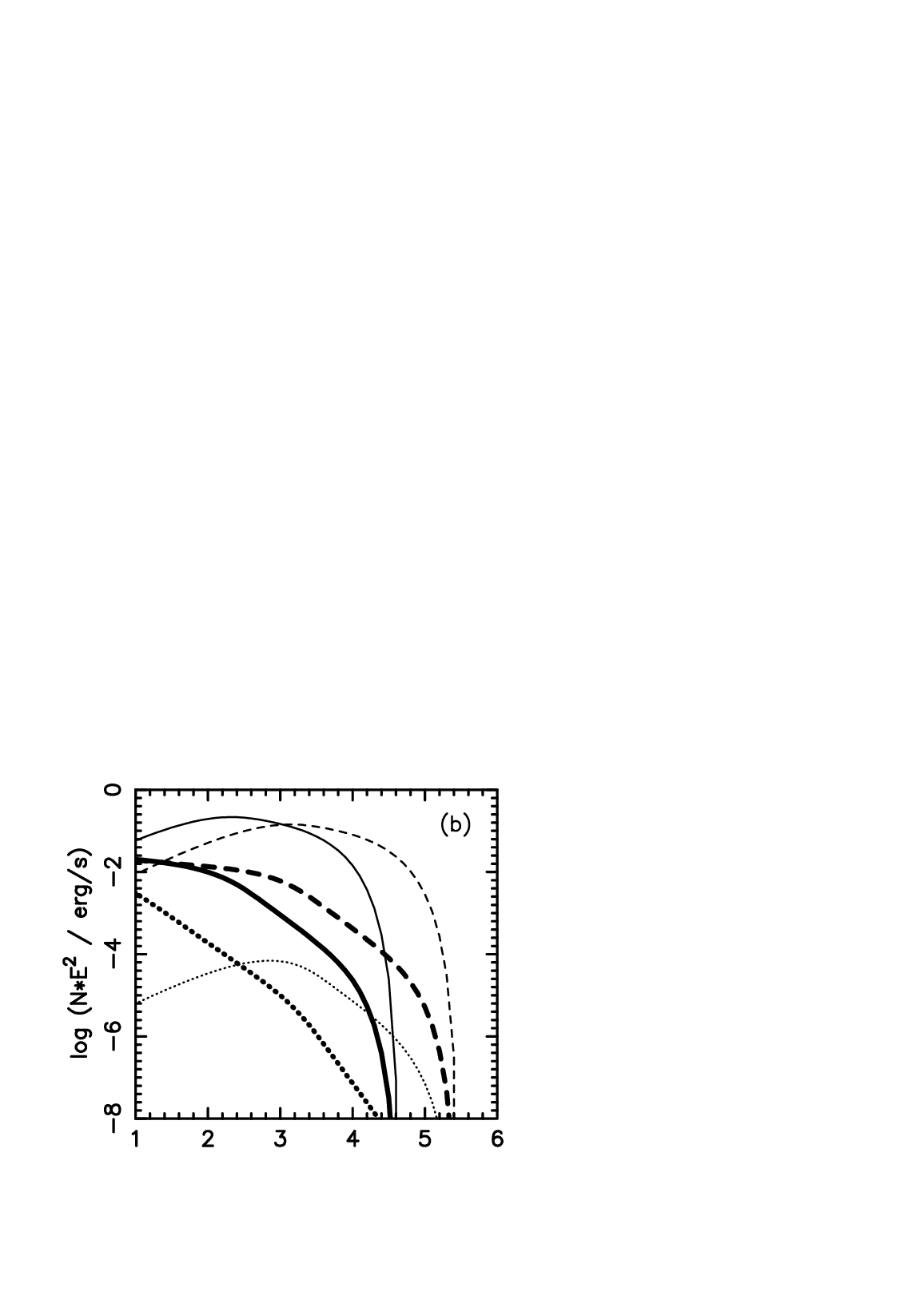}
\includegraphics{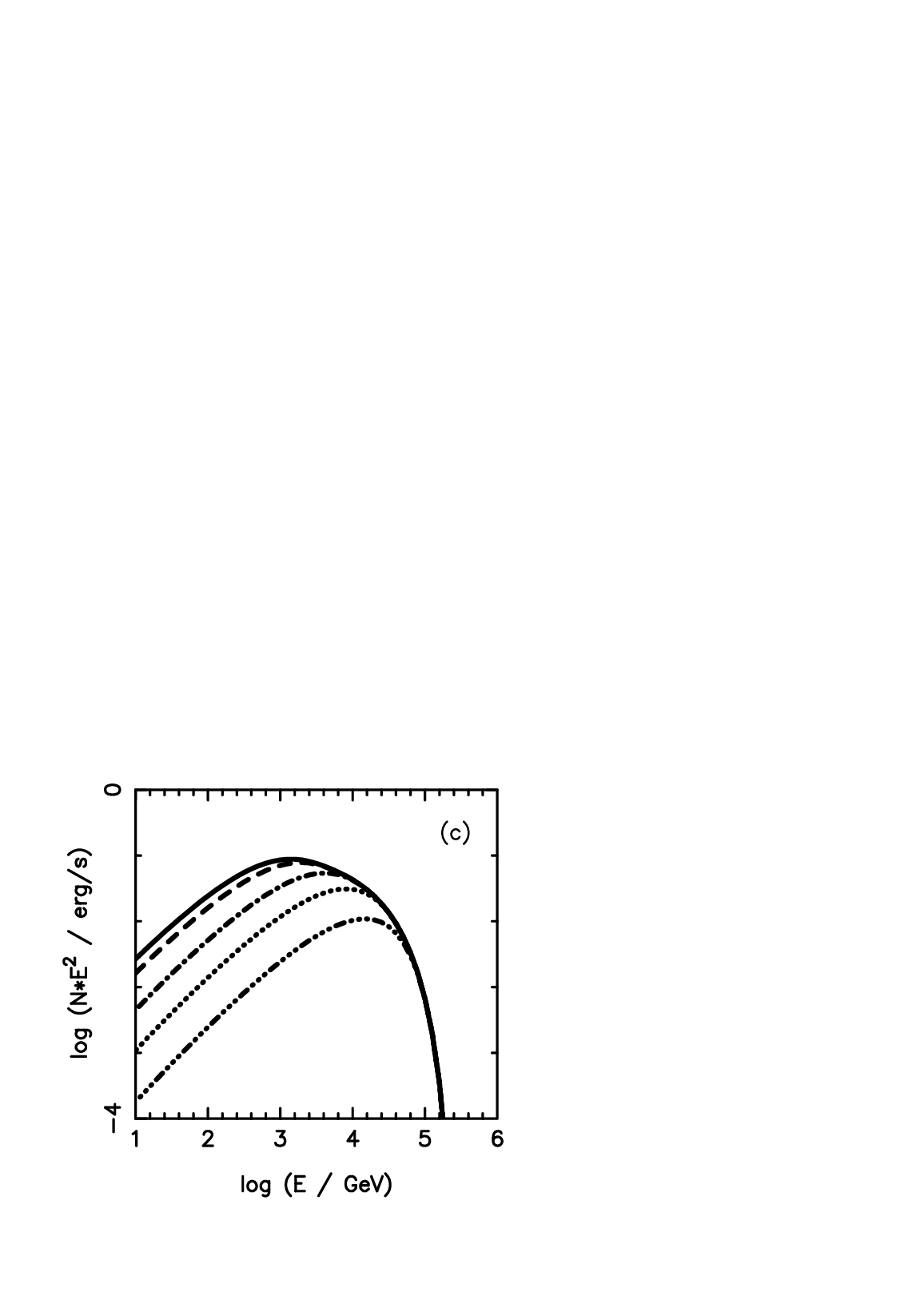}
\caption{$\gamma$-ray spectra (SED) from decay of $\pi^{\rm o}$ produced in the 
interactions of relativistic hadrons with the matter inside the open cluster (thick curves) 
and with dense molecular clouds surrounding the open cluster (thin curves). It is assumed that 
density of matter inside the open cluster is $\rho_{\rm oc}$ = 100 cm$^{-3}$. 
The open cluster is surrounded by dense clouds, with density of $\rho_{\rm cl}$ = $10^4$ cm$^{-3}$, in which all particles escaping from the open cluster are captured (a). 
The results of calculations for $\rho_{\rm oc}$ = 10 cm$^{-3}$ and $\rho_{\rm cl} = 10^5$ 
cm$^{-3}$ are shown in (b). Differential spectrum of hadrons is defined by the high energy cut-off which is due to the diffusion of hadrons from the open cluster (see Eq.~\ref{eq13}), calculated for the magnetic field $B_{\rm oc} = 10^{-5}$ G and $10^{-4}$ G (solid and dashed curves, respectively) and by the spectral index equal to $\alpha = 2.1$. The $\gamma$-ray spectra, produced by hadrons with $\alpha = 3$ and the cut-off for $B = 10^{-5}$ G, are shown by the dotted curves. The comparison of $\gamma$-ray spectra produced in clouds by hadrons escaping from the open cluster are shown under the assumption that only highest energy hadrons with energies in the range, $(0.3-1)\times E_{\rm max}$ (triple dotted-dashed curve), $(0.1-1)\times E_{\rm max}$ (dotted), $(0.03-1)\times E_{\rm max}$ (dot-dashed), $(0.01-1)\times E_{\rm max}$ (dashed), and all escaping (solid), are able penetrate the cloud (figure c).}
\label{fig4}
\end{figure}

We calculate also the $\gamma$-ray spectra produced by hadrons during their propagation
inside the open cluster and after their escape to surrounding molecular clouds, applying the scale break model for hadronic interactions appropriate for hadrons with considered energies (see Wdowczyk \& Wolfendale~1987). As an example, we apply the
parameters of the open cluster Berk 87 and consider two cases with different density of matter inside the open cluster and surrounding clouds (see Fig. \ref{fig4}ab). At first, we assume for simplicity that all
hadrons escaping from the open cluster are captured by the molecular clouds ($\xi = 100\%$). Hadrons are injected with this same rate during the lifetime of the WR star. They obtain the power law spectrum with the high energy cut-off determined by the magnetic field strength inside the cluster. The low energy cut-off is assumed at 10 GeV. The injection rate of hadrons is normalized to 1 erg s$^{-1}$. The $\gamma$-ray spectra
produced by hadrons in their interaction with the matter inside the open cluster show characteristic break at energies determined by their diffusion time from the cluster which in turn depends on the magnetic field strength inside the cluster (typically between $10^2-10^3$ TeV, see thick curves in Fig. \ref{fig4}). Due to relatively low density of matter inside the cluster (in respect to surrounding clouds), relativistic hadrons are not able to interact frequently. Therefore, hadrons with large energies are able to escape from the open cluster. The $\gamma$-ray spectra produced by these hadrons in collisions with the 
matter of the molecular clouds dominate at larger energies (see thin curves). They show a characteristic break at similar energies as the break in the $\gamma$-ray spectra produced inside the open cluster (see thick curves). However, in contrast to hadrons inside the open cluster, relativistic hadrons captured by dense clouds cool efficiently on multiple interactions with the matter already during the lifetime of the WR star due to much larger matter densities. The relative power in the $\gamma$-ray spectra produced inside and outside the open cluster depends on the ratio of density of matter inside these two regions.

We also investigate the $\gamma$-ray production by hadrons assuming that only the most energetic ones are captured inside the molecular clouds (see Fig.~\ref{fig4}c). This might be caused by the energy dependent diffusion of relativistic hadrons from the open clusters. As expected, the $\gamma$-ray
spectra produced by these hadrons are only limited to the highest energy part of the spectrum calculated
in the case of all hadrons captured by the clouds (marked by the solid curve).

\begin{table*}
\caption{The basic parameters of the models.}             
\label{tab2}      
\begin{tabular}{c c c c c c c c c c }     
\hline
Parameter      &  Berk 87  &  Cyg OB2  &  Wester 2 & Berk 87  &  Cyg OB2  &  Wester 2 & Berk 87  &  Cyg OB2  &  Wester 2 \\
\hline
     &  & Model A1 &  &  & Model A2 & &  & Model B &  \\
\hline
magnetic field                             & $10^{-4}$ G         &  $10^{-4}$   &   $10^{-5}$ & $10^{-5}$ G          &  $3\times 10^{-6}$ G   &   $10^{-5}$ G & $10^{-5}$ G     &  $10^{-4}$ G   &  $10^{-4}$ G \\
spectral index (p)                         & 2.1                 &  2.6         &  2.4   & 2.1                 &  2.4 &  2.4   & 2.4                 &  2.4     &    \\
efficiency ($\eta_{\rm p}$)   & $3\times 10^{-3}$   &  $6\times 10^{-3}$  & $5\times 10^{-2}$ & $3\times 10^{-3}$   &  $2\times 10^{-3}$   &  $10^{-2}$  & $10^{-3}$           &  $10^{-3}$      &  $2\times 10^{-3}$ \\  
spectral index (e)      & & &        & & &      & 2.4                 &  2.4  &  2.2 \\
efficiency ($\eta_{\rm e}$) & & & & & &   & $5\times 10^{-3}$   &  $10^{-2}$      &$2\times 10^{-2}$ \\
capturing factor  ($\xi$)      & $10\%$              &  $100\%$      &  $10\%$ & $4\%$               &  $10\%$      &  $6\%$   & $100\%$             &   $100\%$      &  $6\%$ \\              
\hline
\end{tabular}
\end{table*}
\section{Gamma-ray fluxes from specific objects}

We calculate the broad band photon spectra produced by electrons and hadrons inside a few specific open clusters and compare them with the high energy observations. It is assumed that the fluxes (and the upper limits) 
reported at high energies by the EGRET, Cherenkov telescopes (Whipple, HEGRA, HESS), and 
the Milagro detector from direction of these open clusters are real and caused by this same 
single source. These observations are analyzed in terms of two general pictures: model A - all 
the emission above $\sim 100$ MeV due to hadrons, model B - the low energy 
$\gamma$-ray emission (EGRET range) is due to leptons and the high energy emission 
(above $\sim 100$ GeV) is due to hadrons. In all cases it is assumed that the average density
of matter inside the open clusters is 10 cm$^{-3}$ and in the surrounding clouds $10^3$ cm$^{-3}$. 
The parameters of the models applied for these specific sources are collected in Table 2.

\begin{figure}
\vskip 10.2truecm
\includegraphics{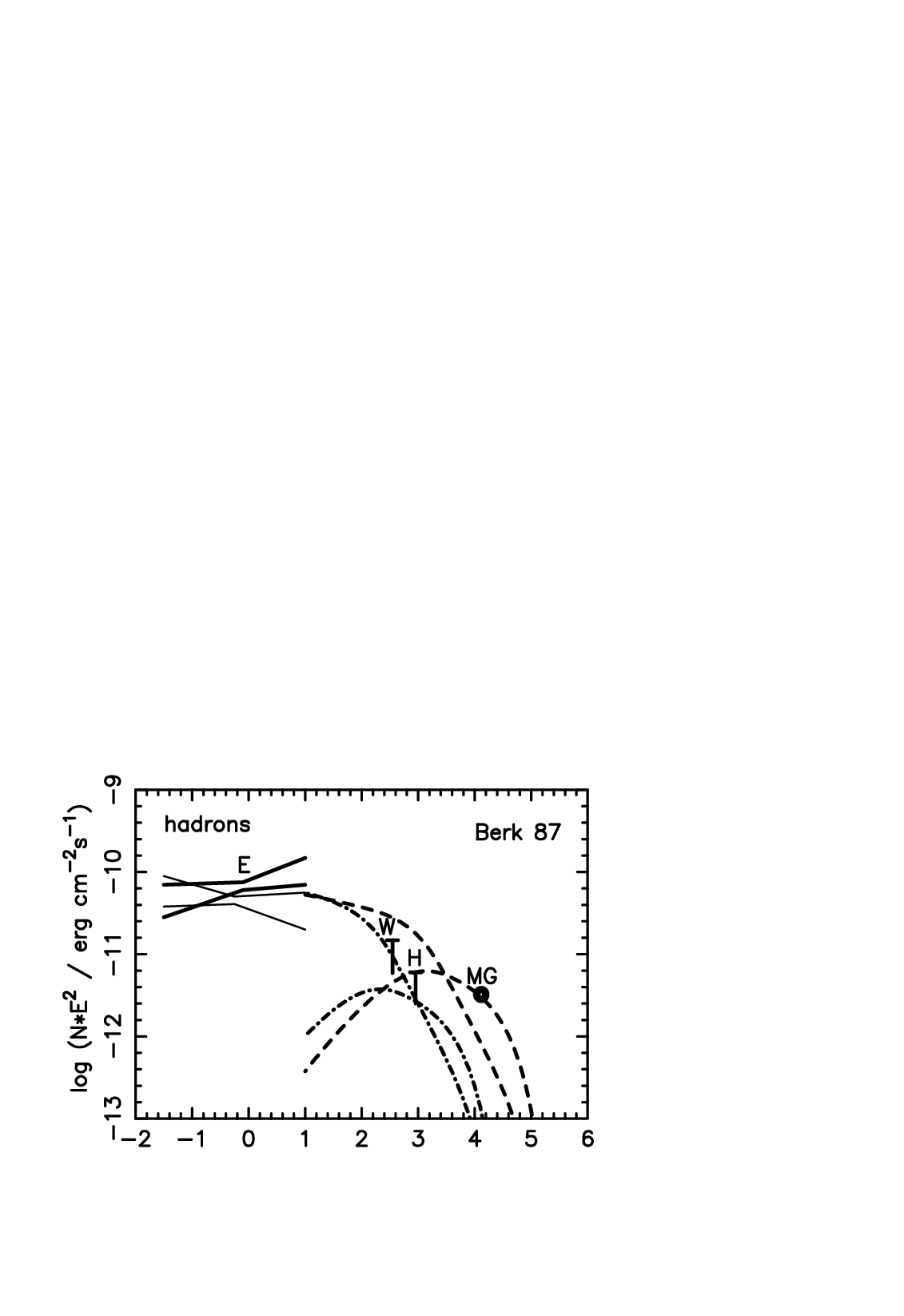}
\includegraphics{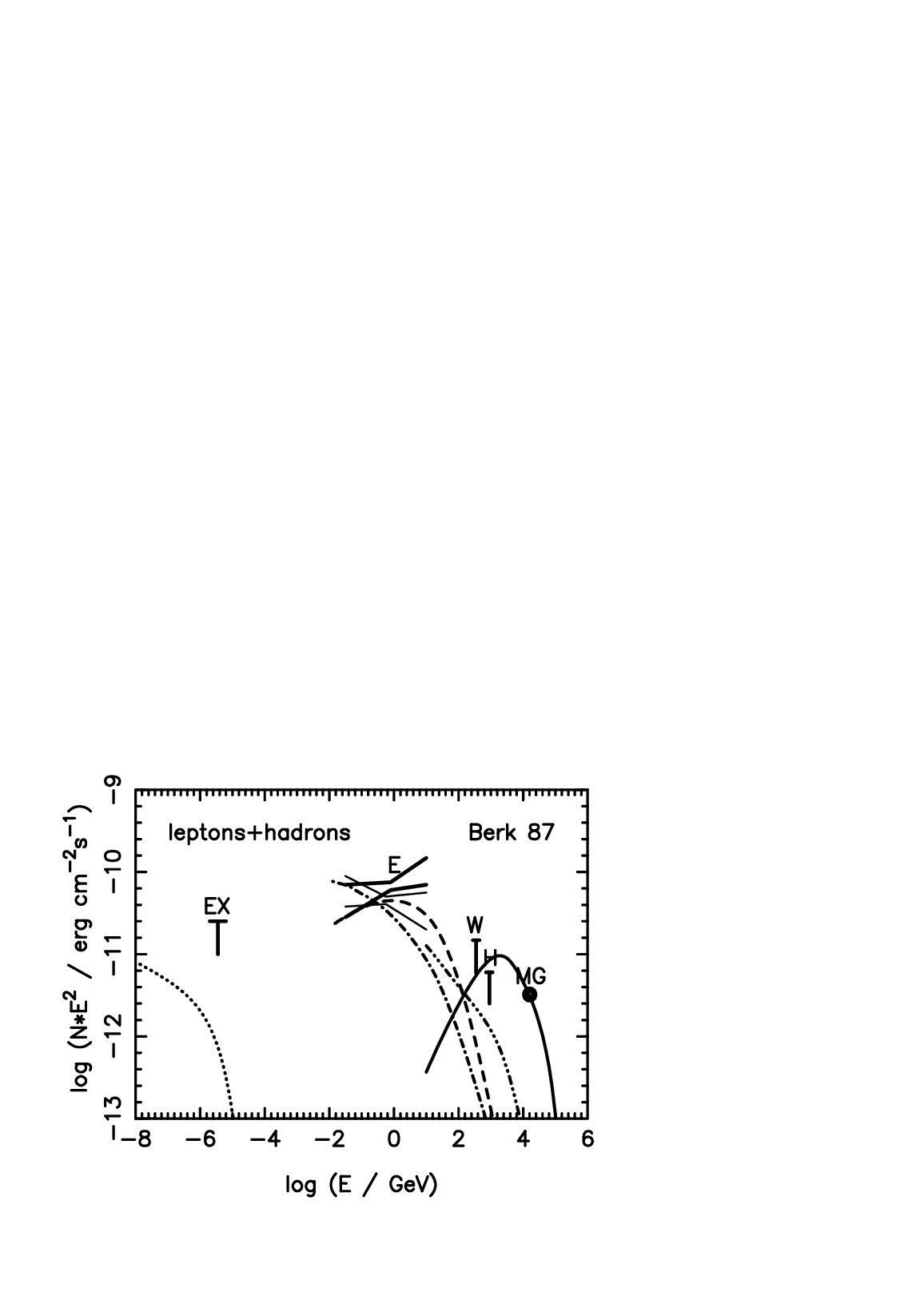}
\caption{The multiwavelength spectrum (SED) observed from the region toward the open cluster 
Berk 87 constrained by the observations with the EXOSAT satellite (marked by EX, 
Warwick et al.~1988), the unidentified EGRET sources from the 3rd Catalog 3EG 2016+3657 
(thin solid) and 3EG J2021+3716 (thick solid) (E, Hartman et al.~1999), the Whipple upper limit 
(W, Fegan et al.~2005), the HEGRA upper limit (H, Aharonian et al.~2006) and the Milagro 
detection of the extended source MGRO J2019+37 (MG, filled circle,  Abdo et al.~2006). 
$\gamma$-ray spectra produced by hadrons interacting with the matter inside the open cluster 
are calculated in terms of the model A (upper figure). Two sets of parameters for this model 
are considered (Table 2, A1 - dashed curves and A2 - dot-dashed in Fig.~5). 
$\gamma$-ray spectra produced in terms of the model B (Table 2)
in the synchrotron (dotted curve), IC (dashed) and bremsstrahlung (dot-dashed) 
processes and the $\gamma$-ray spectra from decay of $\pi^{\rm o}$ produced by hadrons in 
collisions with the matter inisde the open cluster (triple dot-dashed 
curve) and hadrons which escaped from the open cluster into surrounding molecular clouds 
(solid curve). The density of radiation is described by $\mu = 1$.}

\label{fig5}
\end{figure}
\subsection{Berk 87}

The EGRET telescope discovered two point like sources, 3EG J2021+4716 and 3EG J2016+3657, in the 
direction of Berk 87 (Hartman et al.~1999) and the GeV source J2020+3658 (Lamb \& Macomb~1997) 
which is likely related to 3EG J2021+3716 (Roberts et al.~2001). Also diffuse X-ray emission 
from this cluster has been observed by the EXOSAT ($\sim 5\times 10^{32}$ erg s$^{-1}$ for the 
distance of 900 pc, Warwick et al.~1988) and by the ASCA satellites (Roberts et al.~2002). Since this emission can be at least partially thermal, it introduces the upper bound on the 
possible diffuse non-thermal X-ray emission from this open cluster. The region of these EGRET 
sources has been also observed by the Cherenkov telescopes at TeV energies. The Whipple group  
reported only the upper limit on the level of $2\times 10^{-11}$ ph. cm$^{-2} s^{-1}$ above 
350 GeV (Fegan et al.~2005). The HEGRA upper limit at energies above 0.7 TeV on the level of 
$3\times 10^{-13}$ cm$^{-2}$ s$^{-1}$ (based on the 6.4 hr data) has been reported by 
Tluczykont et al.~(2001) and on the level of 
$1.08\times 10^{-12}$ cm$^{-2}$ s$^{-1}$ above 0.9 TeV based on the 13.4 hr data 
(Aharonian et al.~2006). Moreover, very recently the Milagro group reported positive detection 
of a quite extended source at energies $>12$ TeV from the region containing Berk 87 with the 
flux estimated on $1.7\times 10^{-13}$ ph. cm$^{-2}$ s$^{-1}$ (MGRO J2019+37, Abdo et al.~2006). 
All these observational results are collected in Fig.~\ref{fig5}.

By applying the general model for the acceleration of particles and radiation models defined 
as A (pure hadronic) and B (hybrid, hadronic-leptonic), we calculate the expected broad band 
photon emission from the open cluster Berk 87 and its surroundings. The basic parameters 
describing Berk 87 are reported in Table.~1. Let us at first consider in detail model A.
Since the spectra of EGRET sources are flat, we have to assume that hadrons have also  spectral 
index close to 2 with the high energy cut-off defined by the magnetic field 
strength inside the open cluster. $\gamma$-ray spectra from decay of $\pi^o$, produced by 
hadrons inside the open cluster and surrounding dense clouds, are shown for $B = 10^{-5}$ G 
(model A1, dot-dashed curves, in the upper Fig. \ref{fig5}) and $B = 10^{-4}$ G (model A2, dashed curves). 
As we have shown in Sect.~4 (Fig.~\ref{fig4}c) if only the highest energy hadrons are captured inside the clouds, then the highest energy part of the $\gamma$-ray spectrum remains on the level 
expected for the case of complite (energy independent) capturing.
The spectra are on the level of the EGRET flux for the energy conversion
efficiency from the WR star wind (WR 142) to particles shown in Tab.~2. 
Note that the spectrum obtained for weaker magnetic field is not able to explain the extended
emission reported by the Milagro experiment. On the other hand, the $\gamma$-ray spectrum
calculated for stronger magnetic field is clearly above the upper limits reported by the Whipple and HEGRA experiments at TeV energies. 
In the bottom Fig.~\ref{fig5}, we also show the $\gamma$-ray spectra 
produced by  hadrons which are injected with the steeper spectrum (spectral index 2.4) and 
the maximum energies determined by $B = 10^{-4}$ G (inside the open cluster - triple-dot-dashed 
curve, and from dense surrounding clouds - solid curve). 
Note that, $\gamma$-ray emission produced in dense clouds is quite extended, 
and that's why difficult to be observed with the Cherenkov telescopes of the Whipple and HEGRA 
telescopes. These $\gamma$-ray spectra are generally consistent with the observations by different telescopes at the TeV energies provided that spectrum of hadrons has to become significantly flatter 
below $\sim 100$ GeV in order to be consistent with the flat spectrum characteristic for 
the EGRET sources. Such a break in the spectrum is difficult to explain in the simple shock acceleration scenario. Therefore, we conclude that the pure simple hadronic model is unlikely.

In the model B, electrons are assumed to be injected with the power law spectrum and spectral 
index 2.4 extending up to the maximum energies $\sim 10^5$ GeV. The photon spectra 
produced by electrons in the synchrotron, bremsstrahlung and IC processes are shown in 
Fig.~\ref{fig5}. The EGRET spectrum is described in such a model by the combination of 
the bremsstrahlung spectrum (at lower energies) and the IC spectrum (at higher energies).  
Due to the Klein-Nishina effects and efficient colling of electrons with the largest energies on the synchrotron process, the $\gamma$-ray spectra from bremsstrahlung and IC processes have to 
steepen. These spectra do not overcome the upper limits introduced by the Whipple and HEGRA 
telescopes. However, they are not able to explain the extended emission at the highest energies 
reported by Milagro group. In order to explain all these results simultaneously, we postulate 
additional production of $\gamma$-rays by hadrons which have escaped from the open cluster and 
interacted inside surrounding dense clouds. Hadrons are accelerated inside the open cluster 
with the spectral index as postulated for electrons and with the cut-off at energies expected 
for this same value of the magnetic field (i.e. $10^{-5}$ G). The synchrotron spectrum expected 
in such a hybrid model is clearly consistent with the upper limit on the diffuse X-ray emission 
introduced by the EXOSAT data.

\subsection{Cyg OB 2}

\begin{figure}
\vskip 10.2truecm
\includegraphics{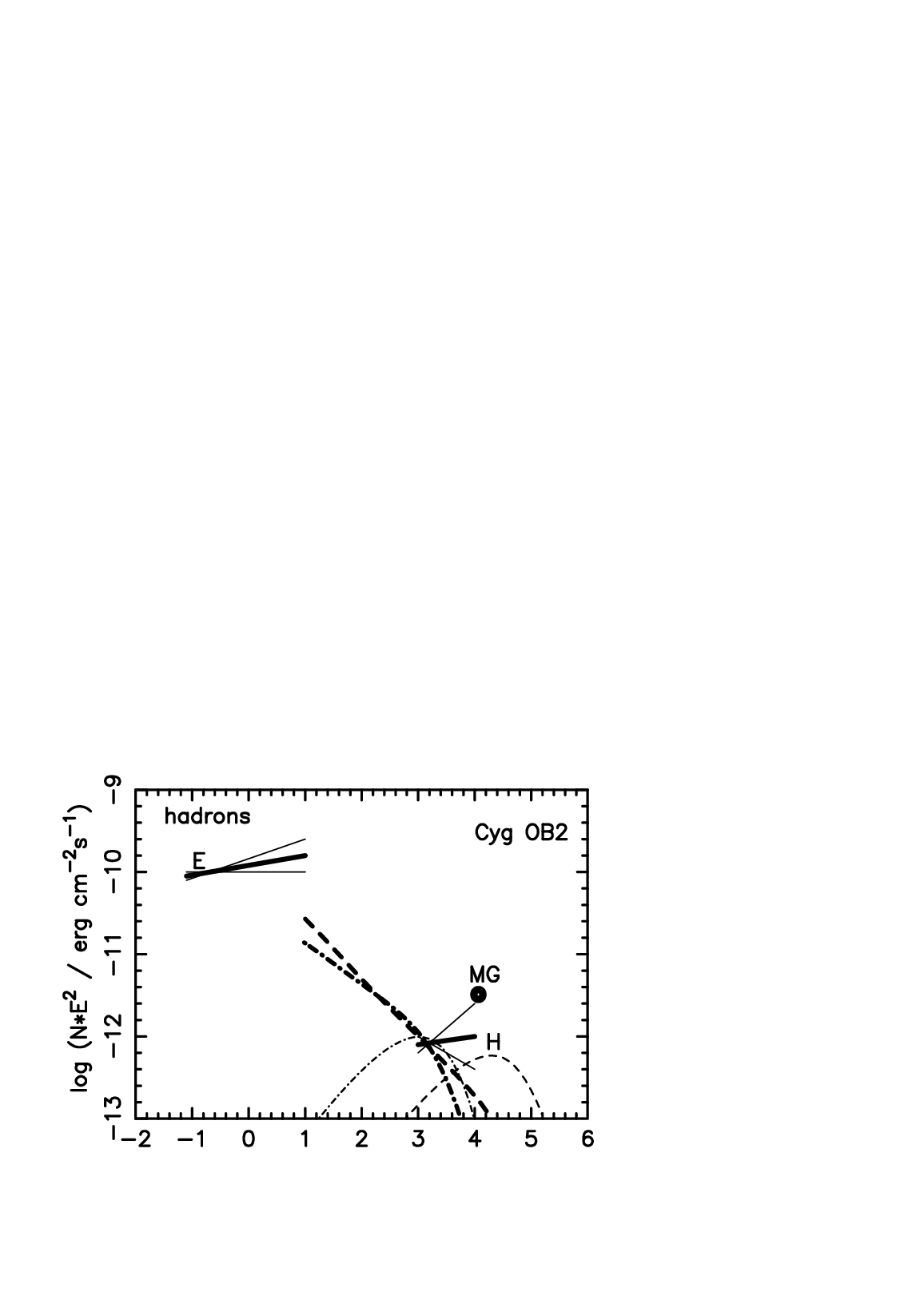}
\includegraphics{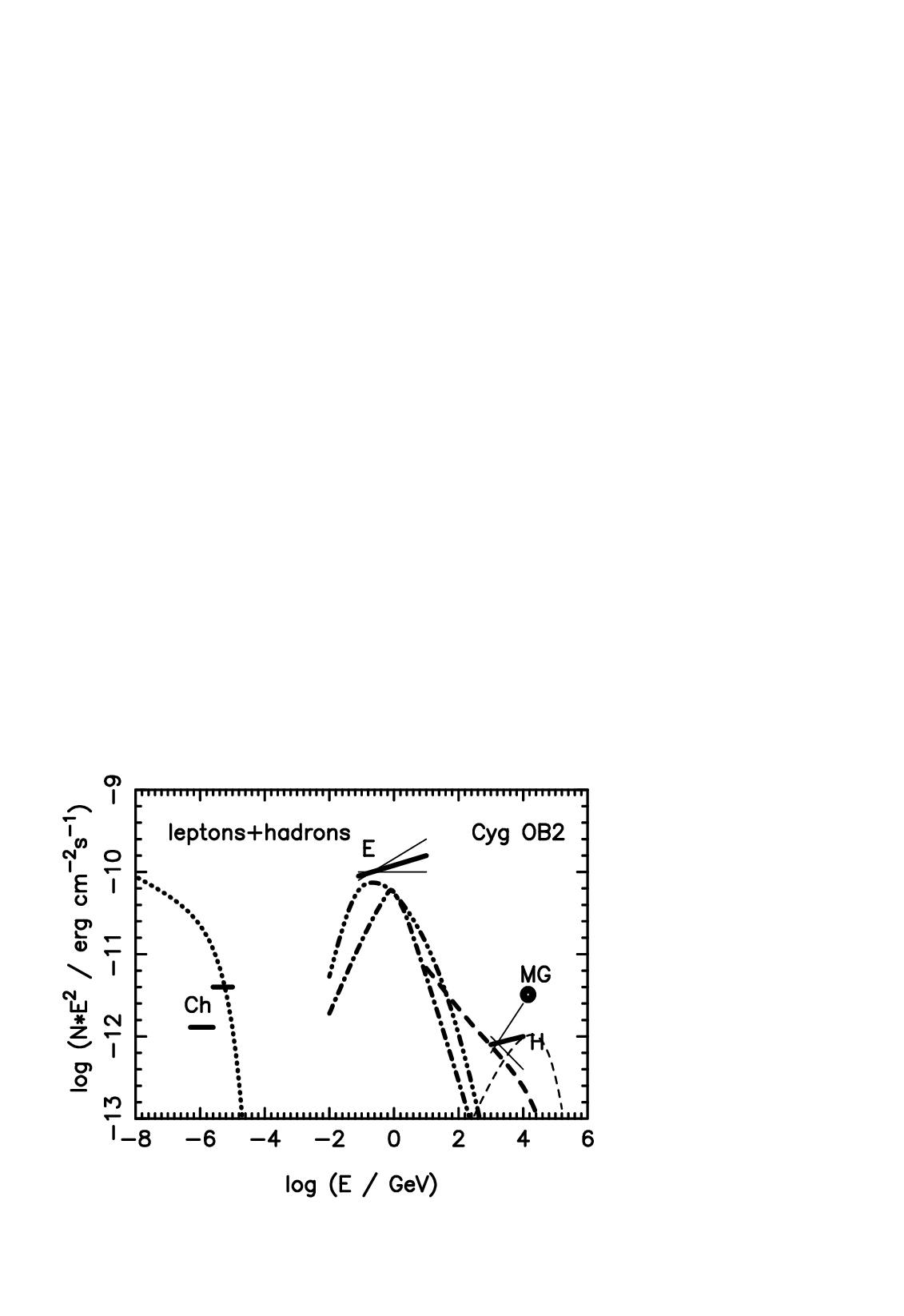}
\caption{As in Fig. \ref{fig5} but for the region toward Cyg OB2. 
The Egret source 3EG J2033+4118 (E, Hartman et al. 1999), the HEGRA source 
(H, Aharonian et al. 2002), the Milagro extended excess (MG, Sinnis et al.~2007), and 
the Chandra upper limits on the diffuse X-ray emission (Ch, Butt et al. 2002). 
The $\gamma$-rays from the open cluster calculated in model A (upper figure) are marked by 
the thick curves and from the surrounding clouds by the thin curves. The comparison with 
the Model B is shown on the bottom figure.
Spectra are produced by leptons in synchrotron process (dotted curve), bremsstrahlung (dot-dashed), 
and IC (triple-dot-dashed), and the $\gamma$-ray spectra (from decay of $\pi^{\rm o}$) are 
produced by hadrons inside the open cluster (thick dashed) and surrounding clouds (thin dashed).}
\label{fig6}
\end{figure}

Between a few $\gamma$-ray sources reported in the direction toward the Cygnus region,
3EG J2033+4118 is believed to be related to the Cyg OB2 (Aharonian et al.~2002, 
Mukherjee et al.~2003, Butt et al.~2003). The spectrum of this  source
was originally described by a simple power law with the index of $1.96\pm 0.10$, 
but a double power law or a power law with the exponential cut-off gives better fits 
(Bertsch et al.~2000, Reimer \& Bertsch~2001). This source is classified as being non-variable 
(Butt et al.~2003). At TeV energies the HEGRA group has reported the discovery of an unidentified, 
steady, TeV source at the edge of the 95$\%$ error circle of the source 3EG J2033+4118 
(Aharonian et al.~2002). The spectrum of this source is also flat (differential spectral index 
$-1.9\pm 0.3_{\rm stat}\pm 0.3_{\rm sys}$), with the flux about two orders below the flux 
observed from 3EG J2033+4118. There is also an indication that the TeV source is extended with 
a radius of $5.6'$ (Aharonian et al.~2002). This detection has been confirmed by the Whipple 
group who reported slightly displaced point like source whose $\gamma$-ray flux is at the level 
of $8\%$ of the Crab signal (Konopelko et al. 2006).
Recently, the Milagro group also reported the $\gamma$-ray flux at energies above 12 TeV from 
the extended region (3x3 square degrees) centered on the HEGRA source equal to 
$\sim 2.4\times 10^{-12}$ cm$^{-2}$ s$^{-1}$. It exceeds by a factor of $\sim 3$ the HEGRA flux.
A weak diffuse X-ray emission has been observed by the ROSAT and Chandra telescopes, 
with the intensity of $1.3\times 10^{-12}$ ergs cm$^{-2}$ s$^{-1}$ in 0.5-2.5 keV and 
$3.6\times 10^{-12}$ ergs cm$^{-2}$ s$^{-1}$ in 2.5-10 keV (Butt et al.~2003). The radio 
observations of this region indicate a weak shell-like radio structure(s) which are at least 
partially non-thermal (Butt et al.~2007). According to the authors, the radio data are compatible 
with one or more young supernova remnants or with large scale cluster shocks induced by 
the violent action of the many massive stars in Cyg OB2.

The $\gamma$-ray spectra are calculated in terms of the model A (Fig.~\ref{fig6})
assuming the parameters of the open cluster and the acceleration process reported in Table 2. 
The HEGRA spectrum extending up to $\sim 10$ TeV can be explained in such a model as 
emission from dense clouds which efficiently capture hadrons escaping from the open cluster. 
Hadrons can also produce lower 
energy $\gamma$-rays inside the open cluster provided that their spectrum is
quite steep (spectral index 2.6). Moreover, the low energy break at $\sim 10-100$ GeV should be 
present in the hadron spectrum in order to obtain consistent description of a relatively flat 
$\gamma$-ray spectrum in the EGRET energy range. As in the case of Berk 87, such a break is 
difficult to explain in the shock acceleration scenario. Therefore, below we also consider 
the hybrid (leptonic-hadronic) model. 

In the hybrid model (B), we explain the EGRET emission as due to leptons and TeV emission as due 
to hadrons. The set of parameters is chosen in such a way to get consistency 
with the observed X-ray and EGRET data. For these same parameters, we calculate also the TeV $\gamma$-ray spectra produced by hadrons inside 
the open cluster and by hadrons captured in surrounding dense clouds. The spectrum of leptons 
has to be rather steep in order not to overcome the synchrotron X-ray flux. 
However, the spectrum of hadrons can not be
too steep in order to produce adequate $\gamma$-ray fluxes at TeV energies. So then, 
available range of parameters of the acceleration scenario and the open cluster is substantially 
limited (Table 2).

\subsection{Westerlund 2}

In the direction of this open cluster at least one 
the EGRET source, 3EG J1027-5817, is clearly visible (the 3rd EGRET catalog, Hartman et al.~1999). 
This source is probably related to 2EG J1021-5835) and GEV J1025-5809 (Lamb \& Macomb~1997). Its spectral 
index is again flat $\alpha = 1.94$ and the average flux is  
$\sim 8\times 10^{-7}$ ph. cm$^{-2}$ s$^{-1}$ above 100 MeV (Hartman et al.~1999). 
The HESS group reported the TeV source in direction of Westerlund 2 (named 
HESS J1023-575). It looks extended with the estimated size of $\sim 0.2$ deg 
(Aharonian et al.~2007). Diffuse X-ray emission has been observed from Westerlund 2 by the ROSAT 
(luminosity $2\times 10^{34}$ erg s$^{-1}$ in the energy range $\sim $1-2.4 keV, 
Belloni \& Mereghetti~1994) and by the Chandra ($3.6\times 10^{34}$ erg s$^{-1}$ in the energy 
range 0.5-8 keV, Townsley et al.~2005) assuming the distance of 8 kpc. This emission is usually 
related to the thermal emission from the shocked winds of the massive stars. Thus, it should be 
rather considered as an upper limit on the possible non-thermal synchrotron X-ray emission. 

\begin{figure}
\vskip 10.2truecm
\includegraphics{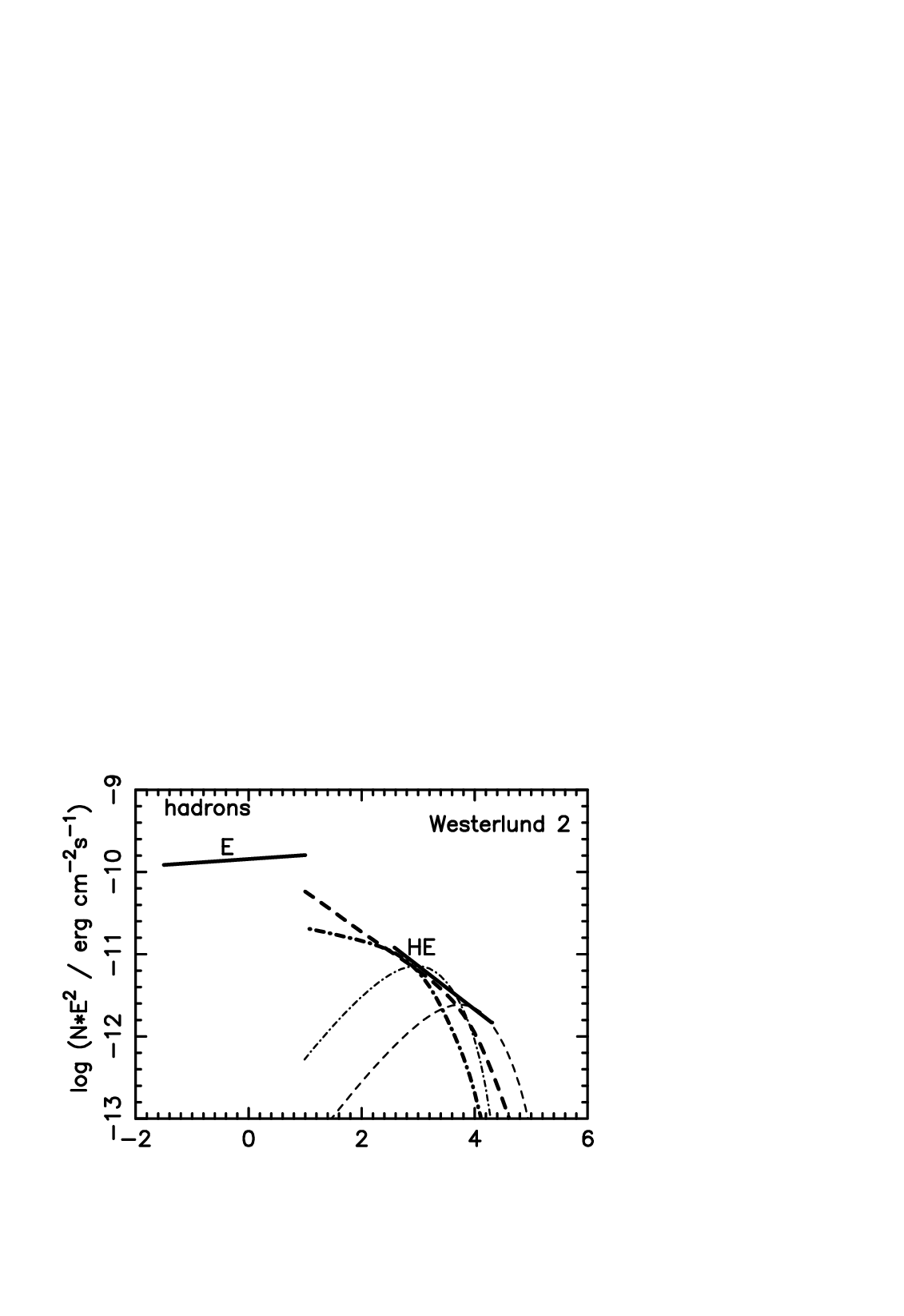}
\includegraphics{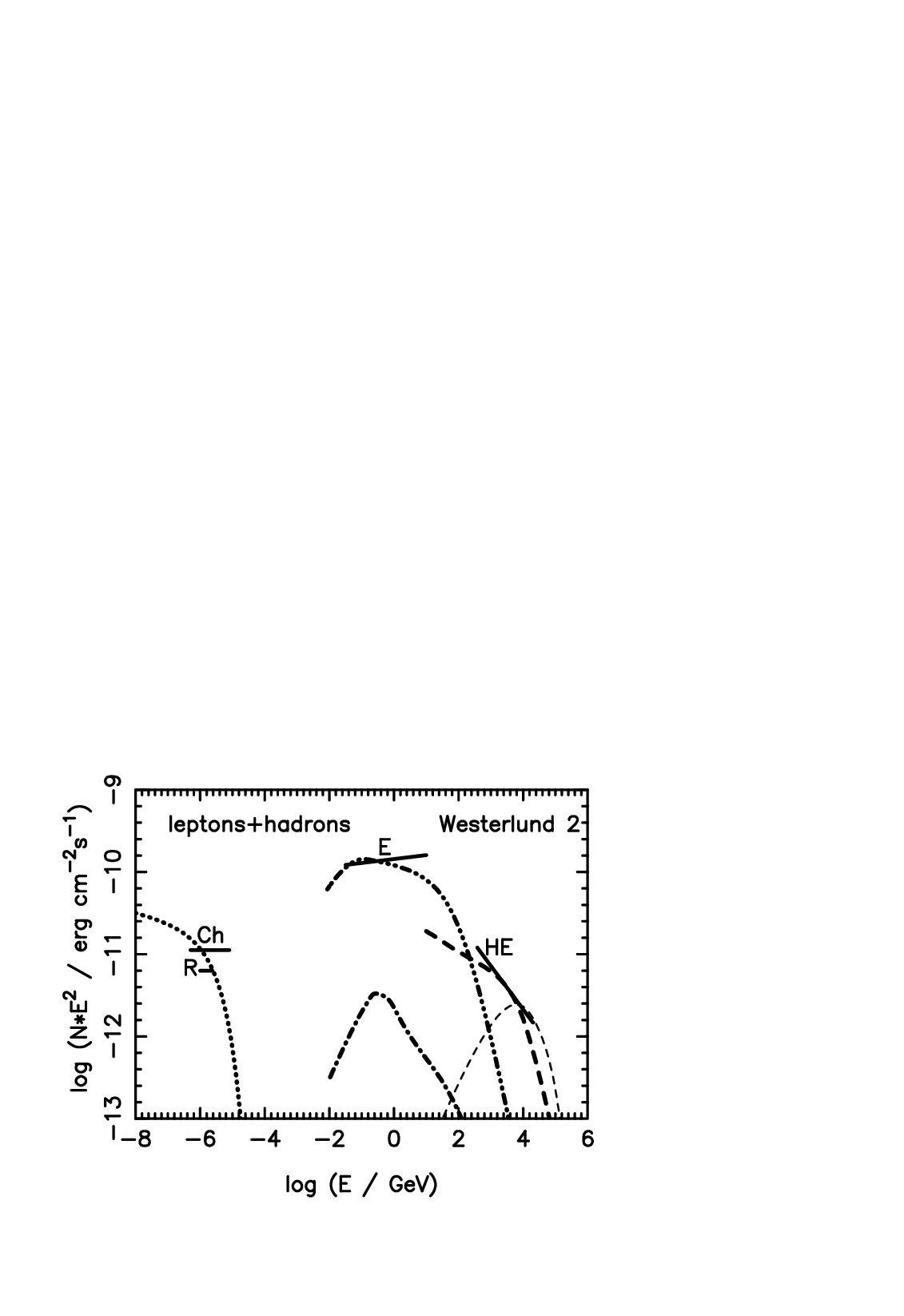}
\caption{As in Fig. \ref{fig6} but for the region toward the open cluster Westerlund 2. 
The Egret source, 3EG J1027-5817 (E, Hartman et al. 1999), the HESS extended source (HE,
Aharonian et al. 2007), the diffuse emission detected by ROSAT (R, Belloni \& Mereghetti~1994) 
and Chandra (Ch, Townsley et al.~2005). The $\gamma$-ray spectra expected in terms of 
the model A (1 and 2) and model B are calculated for the parameters shown in Table 2. 
The radiation density inside the open cluster is defined by $\mu = 10$.}
\label{fig7}
\end{figure}

We have tried to explain at first the $\gamma$-ray fluxes in the GeV-TeV energies
from the direction of Westerlund 2 by a pure hadronic model (model A, see Fig.~\ref{fig7}). 
The HESS spectrum can be fitted as a combination of $\gamma$-rays produced by hadrons inside 
the open cluster and hadrons escaping from the open cluster, and after that captured by 
the molecular clouds. However, 
it is difficult to fit simultaneously the flat spectrum  in the EGRET energy range. 
Hadrons with flat spectrum (spectral index 2.1) predicts too low fluxes at GeV energies. 
On the other hand, hadrons injected with steeper spectrum (spectral index 2.4) predicts 
the GeV emission
inconsistent with the EGRET observations. So then, the explanation of $\gamma$-ray emission 
from Westerlund 2 in the GeV-TeV energy range would require the assumption on the break
in the spectrum of accelerated hadrons at energies $\sim 100$ GeV, which is difficult to
motivate in the simple shock acceleration scenario.

We have also tried to fit the spectra of the EGRET and HESS sources by the pure leptonic model 
with different set of parameters of the open cluster.
However, since electrons lose energy mainly on IC scattering in the strong radiation field of 
Westerlund 2, we were not able to fit the EGRET high level flux and the simple power law 
spectrum reported by HESS in the TeV energies. The reason is that the IC spectrum shows a break 
at TeV energies due to the transition from the Thomson to the Klein-Nishina regime and also 
due to strong synchrotron energy losses of electrons at the higher energy part of their spectrum. 
Therefore, we consider a hybrid electron-hadron model.
In order not to overcome the limit on the diffuse X-ray emission from the open cluster, the spectrum of injected electrons has to be steeper than $\sim$2.2. Therefore, the parameters describing the spectra of particles and the content of the open cluster are already constrained by the available observations (see Table 2). We show the photon spectra produced by electrons and hadrons (model B) in different radiation processes assuming that, 
both types of particles are accelerated with this same spectrum (Fig.~\ref{fig7}). 
Electrons are responsible for the EGRET 
flux (IC scattering), their synchrotron emission is still (marginally) consistent with the 
X-ray observations, and the TeV $\gamma$-rays are produced by hadrons in collisions with 
the matter of the open cluster and surrounding dense clouds. This gives the best description 
of the observations in the broad energy range. The energy density of stellar photons is 
described by the factor $\mu = 10$. This value is much larger than assumed in the case of 
discussed above open clusters but, it is consistent with the expectation that in Westerlund 2, 
only a part of stars present inside the open cluster is taken into account (see Sect.~2.3). 
However, in order to fit simple power law spectrum reported by the HESS experiment, 
the capturing effects of hadrons inside the molecular clouds requires special tuning. 
Otherwise, the $\gamma$-ray spectrum should steepen above $\sim 10$ TeV.

\section{Discussion and Conclusions}

The purpose of this paper is to check if the high energy emission
from some open clusters can be explained with parameterized models
of particle acceleration and radiation in the winds of the WR 
type stars observed at present inside these open clusters. 
Unfortunately, in contrast to e.g. globular clusters, the reach content of the open 
clusters provide interesting target for leptons interacting in different radiation processes.
Therefore, contributions from different processes can become comparable and have to be taken 
into account. This complicates 
significantly theoretical analysis. Moreover, the content and surrounding of the open clusters 
is not very well known. Therefore, open clusters are difficult objects for analysis of details 
of the radiation processes and their eventual comparison with observations. On the other hand, 
recent observations of open clusters in the $\gamma$-ray energy range encourage to pick up 
this interesting astrophysical problem.

Considered model predicts well localized $\gamma$-ray sources at lower energies (e.g. GeV energy 
range) coincident with the open clusters itself, and more extended emission at the highest 
possible TeV energies which should be correlated 
with the surrounding clouds. This is in contrast to the recent 
proposition of Torres at al.~(2004), who postulate production of well localized TeV $\gamma$-ray sources due to interaction of hadrons with dense internal parts of the massive star winds. 
In that model, hadrons accelerated at the massive star winds (or any other unspecified mechanism, i.e. supernova remnants) diffuse against the wind to its dense internal regions. 

Three open clusters have been modeled, Berk 87, Cyg OB2, and Westerlund 2, with
the pure hadronic and hybrid (leptonic-hadronic) models. In principle, pure hadronic models can 
provide good explanation of the whole $\gamma$-ray spectra in the case of Cyg OB2 and 
Westerlund 2, provided that a low energy break in the power law spectrum of hadrons is assumed 
at $\sim 10-100$ GeV. However, such a break is difficult to reconcile in the simple shock 
acceleration scenario. It is even more difficult to fulfill the constraints on the reported 
$\gamma$-ray spectrum from  Berk 87. We were not able to fit simultaneously the EGRET flux, 
the HEGRA and the Whipple upper limits, and the diffuse emission above $\sim 10$ TeV reported by 
the Milagro from the direction of this source.

The situation is much more promising with the hybrid model in which we consider the lower energy
emission as due to leptons (below $\sim 100$ GeV) and the TeV emission as due to hadrons.
For reasonable set of parameters such model can explain flat spectra and high level of the GeV 
emission (EGRET source) and relatively low level of TeV emission which is sometimes 
also characterized by the
flat spectrum (e.g. as in the case of Cyg OB2). In this model, the extended emission above $\sim 10$ TeV
reported from directions of Berk 87 and Cyg OB2, can be explained by hadrons which, diffuse outside the open clusters and, are partially captured in surrounding dense clouds. 
We assumed that leptons and hadrons are injected with these same spectral indices
and the high energy cut-offs determined by this same value of the magnetic field in 
the acceleration region. Note moreover, that these parameters can not be selected arbitrarily.
For example, the observations of the diffuse X-ray flux constrain the allowed value of the spectral 
index of leptons (it can not be too flat). On the other hand, flat spectrum in the GeV energy range 
and the low level of TeV emission constrain the magnetic field strength in the acceleration region.
Unfortunately, the parameters describing the content of the open cluster, surrounding medium,
and the capturing factor of relativistic hadrons by dense clouds are not well determined. 
It is usually argued that density of matter inside the open cluster is moderate due to the 
cleaning effects by the energetic winds of the massive stars. We considered typical values
of the order of 10 particles cm$^{-3}$, in general consistency with derivations by e.g. 
Butt et al.~(2003) 
for the Cyg OB2. Also the effective stellar radiation field seen by leptons during their diffusion inside 
the open cluster is not well determined. We estimated the average density of stellar photons
applying the observed luminosities of the massive stars and the dimension of the open cluster.
But this number can differ significantly from the real effective density seen by leptons due to
the unknown details of their propagation, special distribution of the most luminous stars inside 
the open cluster in respect to the appearance of the shock structures on which these particles 
are accelerated. Due to these inhomogeneities, which are very difficult to take into account, the effective density of stellar photons seen by relativistic leptons can be much larger and their energy losses on ICS process can be significantly enhanced in respect to the synchrotron and bremsstrahlung process.  These different effects makes detailed theoretical interpretation of the high energy processes 
inside the open clusters specially complicated.

Note, that here we only considered the $\gamma$-ray production due to the presence of the observed massive stars inside the specific cluster. The relatively high level of $\sim$10 TeV emission from the most 
massive open cluster Cyg OB2 (reported by the Milagro) can be caused by the contribution from the past WR stars which have already disappeared, or by hadrons accelerated by the supernova shocks 
(e.g. Butt et al.~2007) or relatively young energetic pulsars (e.g. Bednarek 2003). Note, that shell like structures have been discovered inside Cyg OB2. They might be due to the recent supernova explosion.  
Hadrons accelerated by these past evens related to the presence of the massive stars
can be still captured by strong magnetic fields (of the order of $\sim 10^{-3}$ G) at the cores
of the huge molecular clouds surrounding the open clusters. They might contribute 
to the observed large scale diffuse $\gamma$-ray flux at $\sim$10 TeV on a much longer time scale 
than expected in the case of single WR star (i.e. a few $10^5$ yrs).

\section*{Acknowledgments}
This work is supported by the Polish MNiI grant No. 1P03D01028. 


\end{document}